\documentclass[onefignum,onetabnum]{siamonline181217}

\usepackage{lipsum}
\usepackage{graphicx}
\usepackage{amsmath}
\usepackage{amssymb}
\usepackage{setspace}
\usepackage{enumerate}
\usepackage[ruled,linesnumbered,vlined,algo2e,resetcount]{algorithm2e}
\usepackage{color}
\usepackage{array}
\newcolumntype{C}[1]{>{\centering\let\newline\\\arraybackslash\hspace{0pt}}m{#1}}
\usepackage{nicefrac}
\usepackage{microtype}
\usepackage{hyperref}
\usepackage{bm}
\usepackage{multirow}
\usepackage{cite}
\usepackage{microtype}

\definecolor{webgreen}{rgb}{0,.35,0}
\definecolor{webbrown}{rgb}{.6,0,0}
\definecolor{RoyalBlue}{rgb}{0,0,0.9}
\definecolor{purp}{rgb}{0.6,0.05,0.8}
\definecolor{ora}{rgb}{0.7,0.35,0.02}

\newcommand{\p}{\partial}

\newsiamremark{remark}{Remark}
\newsiamremark{hypothesis}{Hypothesis}
\crefname{hypothesis}{Hypothesis}{Hypotheses}
\newsiamthm{claim}{Claim}

\headers{Parallelizable global conformal parameterization}{Gary P. T. Choi, Yusan Leung-Liu, Xianfeng Gu and Lok Ming Lui}

\title{Parallelizable global conformal parameterization of simply-connected surfaces via partial welding\thanks{Submitted to the editors DATE.
\funding{This work was supported in part by the Croucher Foundation (to Gary P. T. Choi) and HKRGC GRF with Project ID: 2130447 (to Lok Ming Lui). }}}

\author{Gary P. T. Choi\thanks{John A. Paulson School of Engineering and Applied Sciences, Harvard University, Cambridge, MA 02138, USA
  (\email{pchoi@g.harvard.edu}).}
\and Yusan Leung-Liu\thanks{Department of Mathematics, The Chinese University of Hong Kong, Hong Kong
  (\email{ylleung@math.cuhk.edu.hk}).}
\and Xianfeng Gu\thanks{Department of Computer Science, Stony Brook University, Stony Brook, NY 11794, USA
  (\email{gu@cs.stonybrook.edu}).}
\and Lok Ming Lui\thanks{Department of Mathematics, The Chinese University of Hong Kong, Hong Kong
  (\email{lmlui@math.cuhk.edu.hk}).}}

\usepackage{amsopn}

\ifpdf
\hypersetup{
  pdftitle={Parallelizable global conformal parameterization of simply-connected surfaces via partial welding},
  pdfauthor={Gary P. T. Choi, Yusan Leung-Liu, Xianfeng Gu and Lok Ming Lui}
}
\fi

\begin{document}

\maketitle

\begin{abstract}

Conformal surface parameterization is useful in graphics, imaging and visualization, with applications to texture mapping, atlas construction, registration, remeshing and so on. With the increasing capability in scanning and storing data, dense 3D surface meshes are common nowadays. While meshes with higher resolution better resemble smooth surfaces, they pose computational difficulties for the existing parameterization algorithms. In this work, we propose a novel parallelizable algorithm for computing the global conformal parameterization of simply-connected surfaces via partial welding maps. A given simply-connected surface is first partitioned into smaller subdomains. The local conformal parameterizations of all subdomains are then computed in parallel. The boundaries of the parameterized subdomains are subsequently integrated consistently using a novel technique called partial welding, which is developed based on conformal welding theory. Finally, by solving the Laplace equation for each subdomain using the updated boundary conditions, we obtain a global conformal parameterization of the given surface, with bijectivity guaranteed by quasi-conformal theory. By including additional shape constraints, our method can be easily extended to achieve disk conformal parameterization for simply-connected open surfaces and spherical conformal parameterization for genus-0 closed surfaces. Experimental results are presented to demonstrate the effectiveness of our proposed algorithm. When compared to the state-of-the-art conformal parameterization methods, our method achieves a significant improvement in both computational time and accuracy.

\end{abstract}



\section{Introduction} 
From mobile games to high-resolution movies, from small 3D printed desk toys to large aircraft engines, 3D geometric models are everywhere nowadays. With the advancement of computing technologies and scanning devices, 3D geometric data can be created or acquired easily. However, at the same time, the scale of the data grows rapidly. In many situations, it is necessary to handle dense geometric data with hundreds of thousands, or even millions of vertices.

In geometry processing, a common representation of 3D objects is triangulated 3D surfaces. To simplify various tasks that are to be performed on the 3D surfaces, one possible way is to transform the 3D surfaces into a simpler 3D shape or a 2D shape. This process is known as surface parameterization. With the aid of surface parameterization, we can perform the tasks on the simpler domain and transform the results back to the original 3D surfaces instead of working on them directly. For instance, under surface parameterization, PDEs on complicated surfaces can be reduced to PDEs on the parameter domain, which are much easier to solve. Also, texture mapping on a 3D surface can be done by parameterizing it onto the 2D plane, in which textures can be easily designed. Among all surface parameterizations, one special type of parameterization is called conformal parameterization, which preserves angle and hence the local geometry of the surfaces. This is particularly important for applications such as texture mapping and remeshing, in which the angle structure plays an important role in the computation. To avoid creating computational burdens or introducing distortions, a fast and accurate method for computing conformal parameterization of surfaces is desired.

\begin{figure}[t]
\centering
\includegraphics[width=0.98\textwidth]{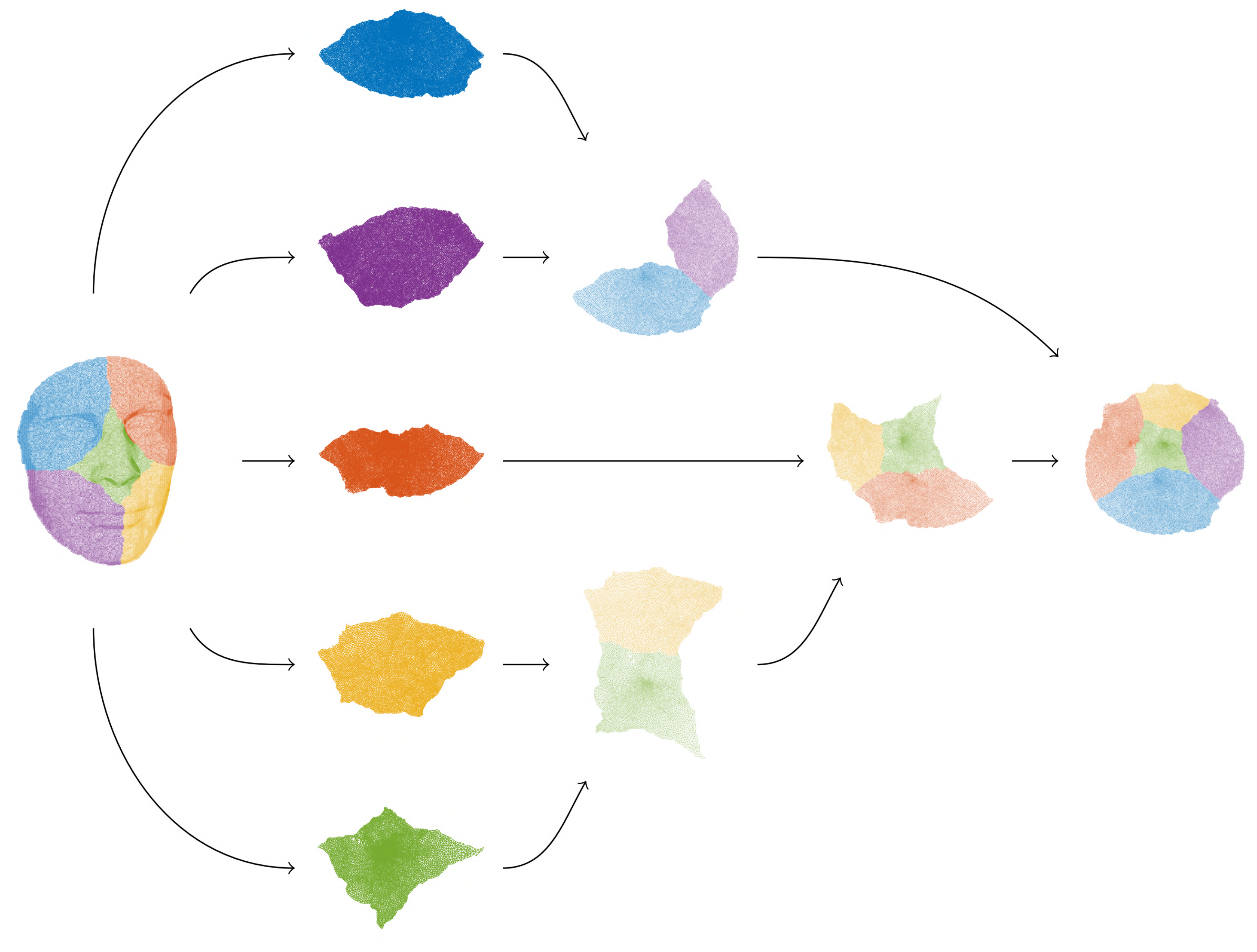}
\caption{An overview of our proposed parallelizable global conformal parameterization (PGCP) algorithm. A simply-connected surface is first partitioned into small subdomains. The subdomains are then conformally flattened onto the plane in parallel. The flattened subdomains are subsequently stitched seamlessly by a novel partial welding technique along the common boundary arcs, thereby producing a global conformal parameterization of the surface. Note that the partial welding step only involves the boundary points of the subdomains but not their interior. The mesh structures of the subdomains are shown only for visualization purpose.}
\label{fig:illustration}
\end{figure}

In this work, we propose a novel \emph{parallelizable global conformal parameterization} method (abbreviated as \emph{PGCP}) for simply-connected surfaces. Unlike the existing methods, our method uses a “divide and conquer” approach and exploits the nature of conformal parameterization, making the computation highly parallelizable. Figure~\ref{fig:illustration} gives an overview of our proposed method. We begin with partitioning a given surface into smaller subdomains. Then, the local conformal parameterizations of the subdomains are computed in parallel. Note that the local parameterization results are not necessarily consistent along their boundaries. Motivated by the theory of conformal welding in complex analysis, we develop a method called \emph{partial welding} to update the boundaries of the flattened subdomains for enforcing the consistency between them. Finally, we solve the Laplace equation with the updated boundary constraints to find conformal parameterizations of the subdomains such that all of them can be glued seamlessly, ultimately forming a global conformal parameterization of the given dense surface. The bijectivity of the parameterization is guaranteed by quasi-conformal theory.

The rest of the paper is organized as follows. In Section~\ref{sect:related}, we review the related works in surface parameterization. In Section~\ref{sect:background}, we introduce the mathematical concepts involved in our work. In Section~\ref{sect:main}, we describe our proposed method for computing a global conformal parameterization of simply-connected surfaces via partial welding. Experimental results and applications are presented in Section~\ref{sect:experiment} for demonstrating the effectiveness of our proposed method. In Section~\ref{sect:discussion}, we discuss the conformality improvement achieved by our method, an extension of our work for reducing the area distortion, and alternative approaches for accelerating the computation. We conclude our work and discuss possible future works in Section~\ref{sect:conclusion}.

\section{Related works}\label{sect:related}
Surface parameterization has been widely studied in geometry processing. For an overview of the subject, readers are referred to the surveys~\cite{Floater05,Sheffer06,Hormann07}. 
It is well-known that only developable surfaces can be isometrically flattened without any distortions in area and angle. For general surfaces, it is unavoidable to introduce distortions in area or angle (or both) under parameterization. This limitation leads to two major classes of surface parameterization algorithms, namely the area-preserving parameterizations and angle-preserving (conformal) parameterizations.

Existing methods for area-preserving parameterizations include the locally authalic map~\cite{Desbrun02}, Lie advection~\cite{Zou11}, optimal mass transport (OMT)~\cite{Zhao13,Su16}, density-equalizing map (DEM)~\cite{Choi18b,Choi18d} and stretch energy minimization (SEM)~\cite{Yueh18}. While the area elements can be preserved under area-preserving parameterizations, the angular distortion is uncontrolled. Since the angular distortion is related to the local geometry of the surfaces, it is important to minimize the angular distortion in many applications such as remeshing, texture mapping and cartography. In those cases, it is preferable to use conformal parameterization. 

Existing conformal parameterization methods for simply-connected open surfaces include the discrete natural conformal parameterization (DNCP)~\cite{Desbrun02}/least-square conformal mapping (LSCM)~\cite{Levy02}, Yamabe flow~\cite{Luo04}, angle-based flattening (ABF)~\cite{Sheffer00,Sheffer05,Zayer07}, circle patterns~\cite{Kharevych06}, spectral conformal mapping (SCP)~\cite{Mullen08}, conformal equivalence of triangle meshes (CETM)~\cite{Springborn08}, discrete Ricci flow~\cite{Jin08,Yang09,Zhang14}, quasi-conformal compositions~\cite{Choi15b,Meng16,Choi17,Choi18a} and conformal energy minimization (CEM)~\cite{Yueh17}. There are also some notable works on the spherical conformal parameterization of genus-0 closed surfaces, including linearization of Laplace equation~\cite{Angenent99,Haker00}, Dirichlet energy minimization~\cite{Gu04}, folding-free global conformal mapping~\cite{Lai14}, FLASH~\cite{Choi15a} and north-south iterative scheme~\cite{Choi16}. More recently, Sawhney and Crane~\cite{Sawhney18} proposed the boundary first flattening (BFF) method, which is capable of computing free-boundary, fixed-boundary and spherical conformal parameterizations for simply-connected surfaces.

Note that all the above-mentioned methods compute a global conformal parameterization of a given surface by handling the entire surface directly. In case the given surface mesh is dense, the computation may be expensive. Also, in case the geometry of the input mesh is complicated, performing a global computation may lead to inaccuracy. Our work aims to overcome these problems by decomposing the input surface mesh into smaller domains and parameterizing them in parallel. The consistency between the domains is ensured by a novel technique called partial welding, thereby forming a global conformal parameterization efficiently.

\section{Mathematical background} \label{sect:background}
\subsection{Harmonic map and conformal map}
Following~\cite{Hutchinson91,Pinkall93}, we introduce the following definitions of harmonic map and conformal map. Let $D$ and $\Omega$ be simply-connected regions in $\mathbb{R}^2$. 

\begin{definition}[Harmonic map]
A map $\varphi:D \to \Omega$ is said to be \emph{harmonic} if it minimizes the Dirichlet energy
\begin{equation}
E_D(\varphi) = \frac{1}{2} \int_D |\nabla \varphi|^2.
\end{equation}
\end{definition}

\begin{definition}[Conformal map]
A map $\varphi:D \to \Omega$ is said to be \emph{conformal} if it satisfies 
\begin{equation}
J \frac{\p \varphi}{\p x} = \frac{\p \varphi}{\p y},
\end{equation}
where $J$ is a rotation by $\frac{\pi}{2}$ in the tangent plane. If we write $\varphi = (\varphi_x, \varphi_y)$, the above equation can be reformulated as the following equations, known as the \emph{Cauchy-Riemann equations}:
\begin{equation}
\left\{
\begin{array}{c}
\frac{\p \varphi_x}{\p x} - \frac{\p \varphi_y}{\p y} = 0, \\
 \frac{\p \varphi_x}{\p y} + \frac{\p \varphi_y}{\p x} = 0.
 \end{array} \right.
\end{equation}
\end{definition}

To achieve conformality, we could minimize the \emph{conformal energy}
\begin{equation}
E_C(\varphi) = \frac{1}{2}\int _D \left[ \left(\frac{\p \varphi_x}{\p x} - \frac{\p \varphi_y}{\p y} \right)^2 + \left(\frac{\p \varphi_x}{\p y} + \frac{\p \varphi_y}{\p x} \right)^2 \right] .
\end{equation}

As shown by Hutchinson~\cite{Hutchinson91}, if we define the \emph{area} $A(\varphi)$ by
\begin{equation}
A(\varphi) = \int_D \left\|\frac{\p \varphi}{\p x} \times \frac{\p \varphi}{\p y}\right\|,
\end{equation}
then the conformal energy can be expressed in terms of the Dirichlet energy and area:
\begin{equation}
E_C(\varphi) = E_D(\varphi) - A(\varphi).
\end{equation}
Since the conformal energy is nonnegative, it follows that the Dirichlet energy is always bounded below by the area. In particular, the equality holds if and only if $\varphi$ is conformal.

Moreover, given the area term $A(\varphi)$, minimizing the conformal energy is equivalent to minimizing the Dirichlet energy. Note that the area depends on how $\varphi$ maps the boundary. In other words, given a ``good'' boundary condition, a conformal map can be obtained by simply finding the harmonic map under the given boundary condition. 

\subsection{M\"obius transformation}
A special type of conformal maps on the extended complex plane $\overline{\mathbb{C}}$ are the \emph{M\"obius transformations}, also known as the \emph{linear fractional transformations}:

\begin{definition}[M\"obius transformation]
A function $f:\overline{\mathbb{C}} \to \overline{\mathbb{C}}$ is said to be \emph{M\"obius transformation} if it is of the form
\begin{equation}
f(z) =  \frac{az+b}{cz+d},
\end{equation}
where $a,b,c,d$ are complex numbers with $ab-bc \neq 0$.
\end{definition}

Given two sets of distinct points $\{z_1,z_2,z_3\}$ and $\{w_1,w_2,w_3\}$, there exists a unique M\"obius transformation satisfying $f(z_i) = w_i$, $i = 1,2,3$. Therefore, M\"obius transformations provides us with a simple way of fixing three points conformally.

\subsection{Conformal welding}
\emph{Conformal welding}, also known as \emph{sewing} or simply \emph{welding}, is a problem in complex analysis which concerns with gluing two surfaces in a conformal way so that they fit together consistently according to certain correspondence. 

\begin{figure}[t]
\centering
\includegraphics[width=0.7\textwidth]{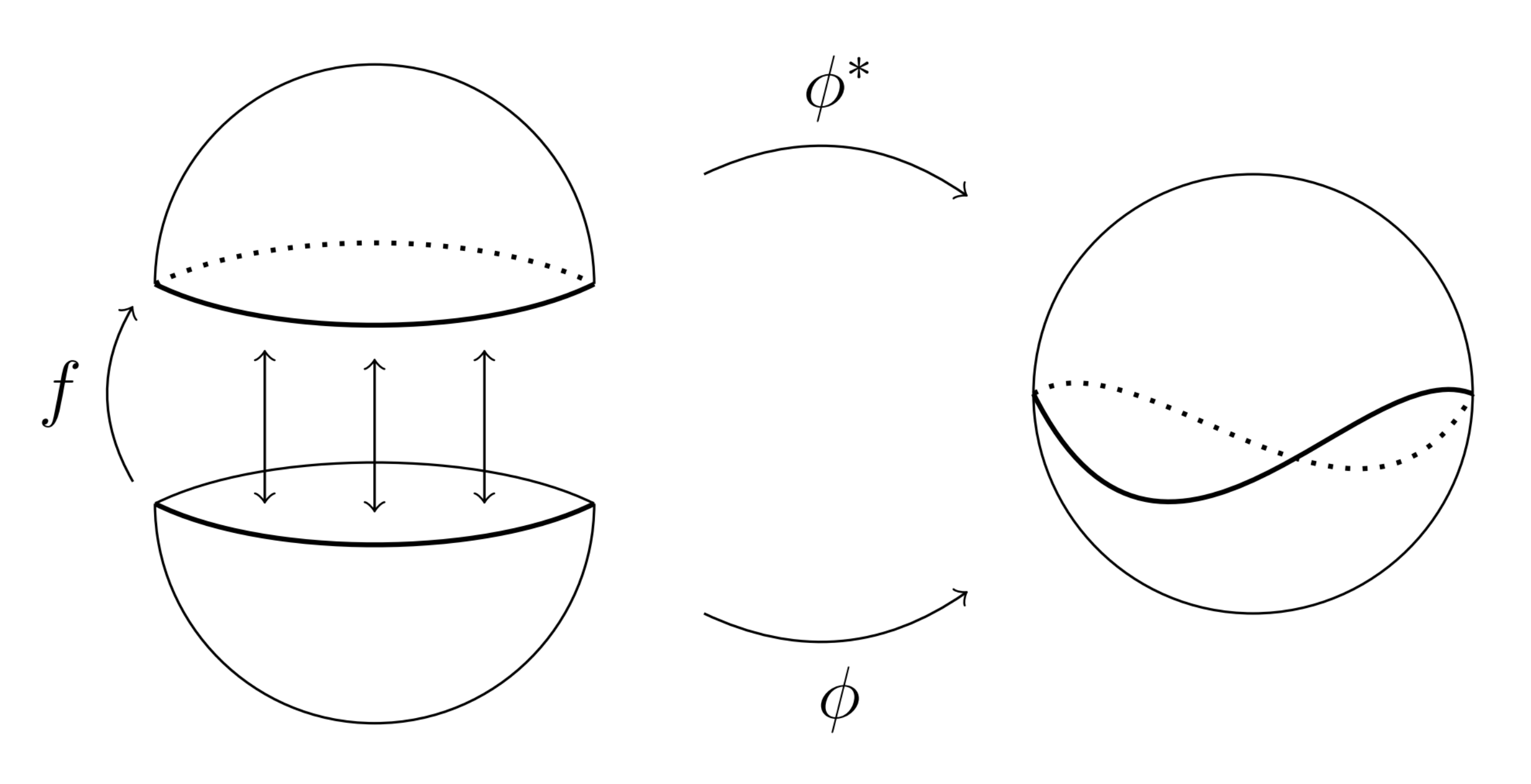}
\caption{An illustration of the closed welding problem. The entire boundaries of the two parts are glued consistently.}
\label{fig:welding_sphere}
\end{figure}

Given a diffeomorphism $f$ from a curve (e.g. the unit circle) to itself, we want to find two Jordan domains $D, \Omega \subset \overline{\mathbb{C}}$ and two conformal maps $\phi:D \to \Omega$ and $\phi^*: D^* \to \Omega^*$ such that $\phi = \phi^* \circ f$ on the curve~\cite{Bishop07}. Here, $D^*$ and $\Omega^*$ are the exterior of $D$ and $\Omega$ respectively. Since $\overline{\mathbb{C}} \cong \mathbb{S}^2$, the two domains $D, \Omega$ can be regarded as two disk-like surfaces on $\mathbb{S}^2$. Intuitively, given a correspondence between the boundaries of the two surfaces, the problem of conformal welding is to find two conformal deformations such that the surfaces are stitched together seamlessly (see Figure~\ref{fig:welding_sphere}). We refer this classical welding problem as a \emph{closed} welding problem. 

For a general homeomorphism $f$, the closed welding problem may not have any solution. However, if $f$ satisfies certain conditions, the problem is solvable. We introduce the concept of \emph{quasisymmetric function} below:
\begin{definition}[Quasisymmetric function~\cite{Lehto73}]
Let $f$ be a continuous, strictly increasing function defined on an interval $I$ of the $x$-axis. We call $f$ $k$-quasisymmetric (or simply quasisymmetric) on $I$ if there exists a positive constant $k$ such that
\begin{equation}
\frac{1}{k} \leq \frac{f(x+t)-f(x)}{f(x) - f(x-t)} \leq k
\end{equation}
for all $x, x-t,x+t \in I$ with $t > 0$.
\end{definition}

One can show that the closed welding problem is solvable if $f$ is a quasisymmetric function from the real axis to itself:
\begin{theorem}[Sewing theorem~\cite{Lehto73}] \label{thm:sewing}
Let $f$ be a quasisymmetric function on the real axis. Then the upper and lower half-planes can be mapped conformally onto disjoint Jordan domains $D, \Omega$ by two maps $\phi, \phi^*$, with $\phi(x) = \phi^*(f(x))$ for all $x \in \mathbb{R}$.
\end{theorem}

The theorem was first proven by Pfluger based on the existence of solutions to the Beltrami equation~\cite{Pfluger60}. Another way to prove the result is to use some approximation techniques on the quasisymmetric function~\cite{Lehto73}.

\subsection{Geodesic algorithm} \label{sect:geodesic_algorithm}
A conformal mapping method called the \emph{zipper algorithm} was proposed independently by K\"uhnau~\cite{Kuhnau83} and Marshall and Morrow~\cite{Marshall87} in the 1980s. In particular, Marshall and Rohde~\cite{Marshall07} proved the convergence of a variant of it called the \emph{geodesic algorithm}. The geodesic algorithm computes a conformal map from a region in the complex plane to the upper half-plane $\mathbb{H}$. Below, we briefly describe the geodesic algorithm.

\begin{figure}[t]
\centering
\includegraphics[width=0.95\textwidth]{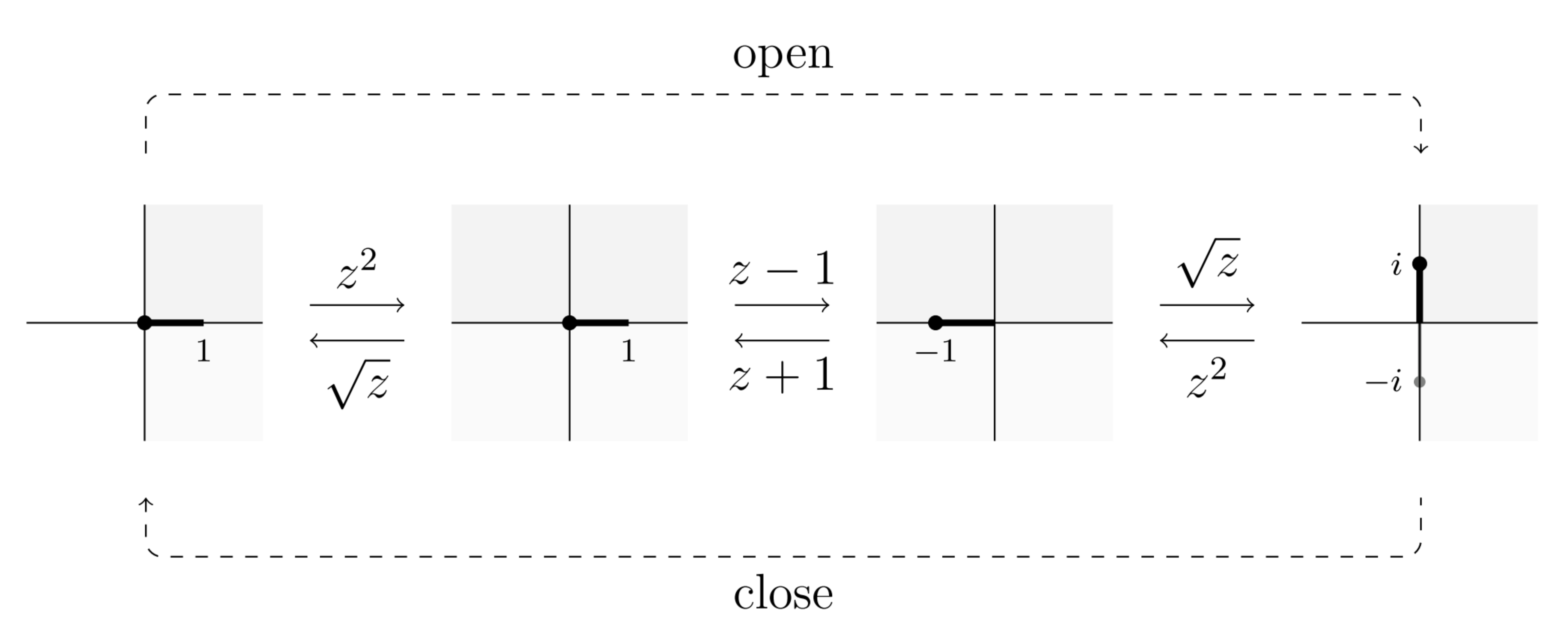}
\caption{An illustration of the opening map and the closing map. The top part shows the processes involved in the opening map $z \mapsto \sqrt{z^2-1}$, which ultimately map $\{0,1\}$ to $\{i,0\}$ or $\{-i,0\}$, depending on the choice of branching. The closing map $z \mapsto \sqrt{z^2+1}$ reverses the processes such that $\pm i$ will be mapped back to 0.}
\label{fig:open_close}
\end{figure}

\begin{figure}[t]
\centering
\includegraphics[width=0.8\textwidth]{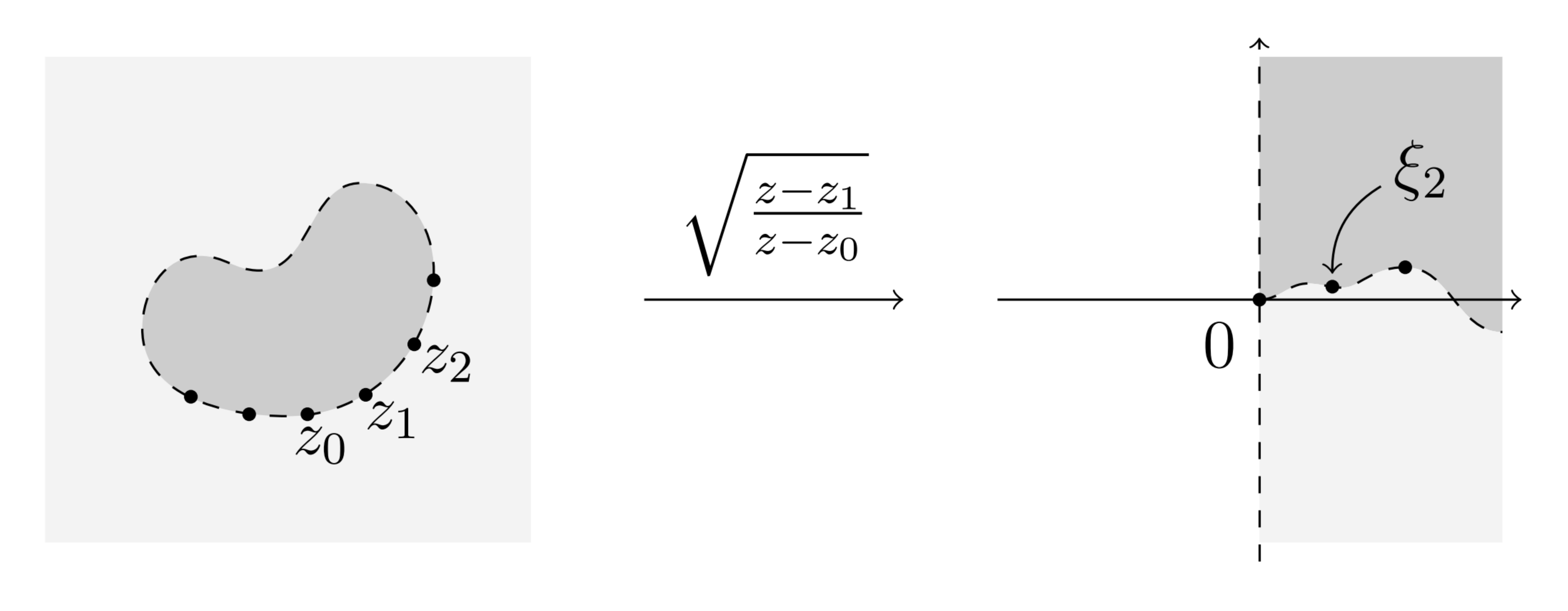}
\caption{In the geodesic algorithm, the first map $g_1$ maps $z_0$ to $\infty$ and $z_1$ to 0. It initiates the process so that we can apply the opening maps for the remaining data points.}
\label{fig:g1}
\end{figure}

The key ingredients of the geodesic algorithm are two maps: the \emph{opening map} $z \mapsto \sqrt{z^2-1}$ and the \emph{closing map} $z \mapsto \sqrt{z^2+1}$. Intuitively, they are operations analogous to opening and closing a slit, behaving like a zipper (see Figure~\ref{fig:open_close}). Suppose we have a simple closed region $\Omega$, and a sequence of boundary points $\{z_0, z_1, \dots, z_k\}$ on $\p \Omega$. To initiate the process, define a map $g_1:\Omega \to \mathbb{C}$ by
\begin{equation}
g_1(z) = \sqrt{\frac{z- z_1}{z-z_0}},
\end{equation}
assuming the branching $(-1)^{1/2} = i$. This maps $\Omega$ to the right half-plane. In particular, the line segment between $z_0$ and $z_1$ is mapped onto the imaginary axis, with $z_0$ mapped to $\infty$ and $z_1$ mapped to 0 (see Figure~\ref{fig:g1}).

Analogously, one can define a map $g_2$ such that the line segment between $g_1(z_1) = 0$ and $\xi_2:=g_1(z_2)$ is mapped to the imaginary axis, while the remaining points are still in the right half-plane. By further repeating the above process, all the boundary points can be pushed onto the imaginary axis one by one. More explicitly, suppose the point $z_j$ has already been transformed to the position $\xi_j$ after applying the opening maps $g_1, g_2, \dots, g_{j-1}$, i.e.
\begin{equation}
\xi_j = g_{j-1} \circ g_{j-2} \circ \cdots \circ g_1(z_j).
\end{equation}
Consider a M\"obius transformation 
\begin{equation}
L_{\xi_j}(z) := \frac{\frac{\text{Re}(\xi_j)}{|\xi_j|^2}z}{1+\frac{\text{Im}(\xi_j)}{|\xi_j|^2}zi}.
\end{equation}
It can be easily checked that $L_{\xi_j}$ maps $\{0, \xi_j, \rho\}$ to $\{0, 1, \infty\}$, where $\rho$ is a point on the imaginary axis at which the orthogonal circular arc from $0$ to $\xi_k$ extends to. Now, as $L_{\xi_j}(\xi_j) = 1$, we can map the segment between $L_{\xi_j}(g_{j-1} \circ g_{j-2} \circ \cdots \circ g_1(z_{j-1})) = 0$ and $L_{\xi_j}(\xi_{j}) = 1$ onto the imaginary axis as illustrated in Figure~\ref{fig:open_close}. Define $g_j$ as the composition of $L_{\xi_j}$ with the opening map $f(z) = \sqrt{z^2-1}$:
\begin{equation}
g_j(z) := \sqrt{L_{\xi_j}(z)^2 -1}.
\end{equation}
Note that $g_j((g_{j-1} \circ \cdots \circ g_1(z_{j-1})) = \sqrt{0-1} = i$ (assuming the branching $(-1)^{1/2} = i$), $g_j(\xi_j) = \sqrt{L_{\xi_j}(\xi_j)^2 -1} = \sqrt{1-1} = 0$, and the entire region will remain in the right half-plane. Therefore, the requirements for $g_j$ are satisfied.

After obtaining the maps $g_1, g_2, \dots, g_k$ such that $g_k \circ g_{k-1} \circ \cdots \circ g_1$ maps all boundary points $\{z_0, z_1, \dots, z_k\}$ onto the imaginary axis, define a final map
\begin{equation}\label{eqt:g_final_geodesic}
g_{k+1}(z) = \left(\frac{z}{1-\frac{z}{g_k \circ g_{k-1} \circ \cdots \circ g_1 (z_0)}}\right)^2.
\end{equation}
$g_{k+1}$ maps the transformed region $g_k \circ g_{k-1} \circ \cdots \circ g_1(\Omega)$ onto the upper half-plane $\mathbb{H}$. Since all the above maps are analytic and, in particular, a square and a square root map are applied in each step, the composition map $(g_{k+1} \circ g_k \circ \cdots \circ g_1): \Omega \to \mathbb{H}$ is conformal. 

\subsection{Quasi-conformal map} \label{sect:qc}
Quasi-conformal map is an extension of conformal map in the sense that it allows for bounded conformal distortion. Intuitively, conformal maps map infinitesimal circles to infinitesimal circles , while quasi-conformal maps map infinitesimal circles to infinitesimal ellipses with bounded eccentricity. The formal definition of quasi-conformal map is given below.

\begin{definition}[Quasi-conformal map~\cite{Gardiner00}]
A map $\varphi:D \to \Omega$ is said to be \emph{quasi-conformal} if it satisfies the \emph{Beltrami equation}
\begin{equation}
\frac{\p \varphi}{\p \bar{z}} = \mu_{\varphi}(z) \frac{\p \varphi}{\p z}
\end{equation}
for some complex-valued function $\mu_{\varphi}$ with $\|\mu_{\varphi}\|_{\infty} < 1$. $\mu_{\varphi}$ is said to be the \emph{Beltrami coefficient} of $\varphi$.
\end{definition}

The Beltrami coefficient $\mu_{\varphi}$ captures the conformal distortion of $\varphi$. In particular, if $\mu_{\varphi} = 0$, then the Beltrami equation becomes the Cauchy-Riemann equations and hence $\varphi$ is conformal. Also, the Jacobian $J_{\varphi}$ of $\varphi$ is given by
\begin{equation}
J_{\varphi} = \left|\frac{\p \varphi}{\p z} \right|^2 (1-|\mu_{\varphi}|^2).
\end{equation}
Therefore, a map is folding-free if and only if its Beltrami coefficient is with sup norm less than 1.

Moreover, one can correct the conformal distortion and non-bijectivity of a map by composing it with another map. If $\varphi_1:\mathbb{C} \to \mathbb{C}$ and $\varphi_2:\mathbb{C} \to \mathbb{C}$ are two maps with Beltrami coefficients $\mu_{\varphi_1}$ and $\mu_{\varphi_2}$, then $\varphi_2 \circ \varphi_1$ is a quasi-conformal map with Beltrami coefficient
\begin{equation}\label{eqt:composition}
\mu_{\varphi_2 \circ \varphi_1}(z) = \frac{\mu_{\varphi_1}(z) +\frac{\overline{{\varphi_1}_z}}{{\varphi_1}_z} \mu_{\varphi_2}(\varphi_1(z))}{1+\frac{\overline{{\varphi_1}_z}}{{\varphi_1}_z} \overline{\mu_{\varphi_1}(z)}\mu_{\varphi_2}({\varphi_1}(z))}.
\end{equation}
In particular, if $\mu_{\varphi_2} = \mu_{\varphi_1^{-1}}$, then $\mu_{\varphi_2 \circ \varphi_1} = 0$ and hence the composition map $\varphi_2 \circ \varphi_1$ is conformal and folding-free. This idea of quasi-conformal composition has been used in~\cite{Choi15a,Choi17,Choi18a}, and the details of the theory and computation of it can be found therein.

\section{Proposed method}\label{sect:main}
Let $\mathcal{S}$ be a simply-connected surface in $\mathbb{R}^3$, with a triangle mesh representation $(\mathcal{V},\mathcal{F})$ where $\mathcal{V}$ is the vertex set and $\mathcal{F}$ is the face set. Our goal is to compute a global conformal parameterization of $\mathcal{S} = (\mathcal{V},\mathcal{F})$ in an efficient and accurate way.

\subsection{Surface partition}
The first step is to partition $\mathcal{S}$ into submeshes based on a prescribed set of edges $\tilde{\mathcal{E}}$. Here, $\tilde{\mathcal{E}}$ can either be defined manually by the user or computed automatically using some existing partitioning methods. This allows the user to have full control of the number of subdomains to be used and how the surface is to be partitioned. More specifically, denote the edge set of $\mathcal{S}$ by $\mathcal{E}$, and the set of boundary edges of $\mathcal{S}$ by $\mathcal{E}_{\text{bdy}}$. Consider the set $E = \mathcal{E} \setminus \left(\tilde{\mathcal{E}} \cup \mathcal{E}_{\text{bdy}} \right)$. We construct a graph $G$ using $E$ and find all connected components in $G$. Suppose there are $K$ connected components in $G$, where each of them consists of a sub-face set $\mathcal{F}_i$, $i = 1, \dots, K$. By tracking all vertices that are contained in $\mathcal{F}_i$, we obtain a sub-vertex set $\mathcal{V}_i$. In other words, we have obtained $K$ simply-connected open submeshes $\mathcal{S}_1 = (\mathcal{V}_1,\mathcal{F}_1), \mathcal{S}_2 = (\mathcal{V}_2,\mathcal{F}_2), \dots, \mathcal{S}_K = (\mathcal{V}_k,\mathcal{F}_K)$ that satisfy the following properties:
\begin{enumerate}[(i)]
\item The union of the vertex sets of all submeshes is exactly $\mathcal{V}$:
\begin{equation}
\bigcup_{i=1}^{K} \mathcal{V}_i = \mathcal{V}.
\end{equation} 
\item The union of the face sets of all submeshes is exactly $\mathcal{F}$:
\begin{equation}
\bigcup_{i=1}^{K} \mathcal{F}_i = \mathcal{F}.
\end{equation} 
\item The intersection of any two different sub-vertex sets is the intersection of the boundary sets, which is either an empty set or a boundary segment:
\begin{equation}
\mathcal{V}_i \cap \mathcal{V}_j = \p \mathcal{S}_i \cap \p \mathcal{S}_j \text{ for all } i,j.
\end{equation} 
\item The intersection of any two different sub-face sets is empty:
\begin{equation}
\mathcal{F}_i \cap \mathcal{F}_j = \emptyset \text{ for all } i \neq j.
\end{equation} 
\end{enumerate}

\subsection{Local conformal parameterization of submeshes}
The next step is to compute a conformal parameterization of every $\mathcal{S}_i$. To find a conformal parameterization $\varphi_i: \mathcal{S}_i \to \mathbb{R}^2$, the DNCP method~\cite{Desbrun02} is used. In short, DNCP minimizes the Dirichlet energy $E_D(\varphi)$ and maximizes the area $A(\varphi)$, based on the fact that the Dirichlet energy is bounded below by the area and conformality is attained when equality holds. We briefly describe the method below.

Let $\mathcal{V}_i = \{v_{i_1}, v_{i_2}, \dots, v_{i_{n_i}}\}$ be the vertices in $\mathcal{S}_i$, and $\varphi_i: \mathcal{S}_i \to \mathbb{R}^2$ be a flattening map. Denote ${\bf u} = \left[{\bf u}_1, {\bf u}_2, \dots, {\bf u}_{n_i}\right]^t = \left[\varphi_i(v_{i_1}), \varphi_i(v_{i_2}), \dots, \varphi_i(v_{i_{n_i}})\right]^t$. The Dirichlet energy is discretized using the cotangent formula~\cite{Pinkall93}:
\begin{equation}
E_D({\bf u}) = \frac{1}{2} \sum_{(v_{i_p},v_{i_q}) \text{ adjacent}} (\cot \alpha_{pq} + \cot \beta_{pq}) |{\bf u}_p - {\bf u}_q|^2 = {\bf u}^t L^{\text{cotan}} {\bf u},
\end{equation}
where $\alpha_{pq}, \beta_{pq}$ are the two angles opposite to the edge $[v_{i_p},v_{i_q}]$ in $\mathcal{S}_i$, and $L^{\text{cotan}}$ is a $|\mathcal{V}_i| \times |\mathcal{V}_i|$ sparse symmetric positive definite matrix also known as the cotangent Laplacian:
\begin{equation}
L^{\text{cotan}}_{p,q} = \left\{\begin{array}{ll}
\frac{1}{2}(\cot \alpha_{pq} + \cot \beta_{pq}) & \text{ if } (v_{i_p},v_{i_q}) \text{ are adjacent},\\
- \sum_{r \neq p} L^{\text{cotan}}_{p,r} & \text{ if } p = q,\\
0 & \text{ otherwise.}
\end{array}\right.
\end{equation}

The area is discretized using the boundary vertices of $\mathcal{S}_i$:
\begin{equation}
A({\bf u}) = \frac{1}{2} \sum_{[v_{i_p},v_{i_q}] \subset \p \mathcal{S}_i} (x_p y_q - y_p x_q) = \begin{pmatrix} {\bf x}^t & {\bf y}^t \end{pmatrix} M^{\text{area}} \begin{pmatrix} {\bf x} \\ {\bf y} \end{pmatrix},
\end{equation}
where ${\bf u}_j = (x_j, y_j)$ for all $j$, ${\bf x} = \left[x_1, x_2, \dots, x_{n_i}\right]^t$ is the collection of all $x$-coordinates of ${\bf u}$, ${\bf x} = \left[y_1, y_2, \dots, y_{n_i}\right]^t$ is the collection of all $y$-coordinates of ${\bf u}$, and $M^{\text{area}}$ is a $2|\mathcal{V}_i| \times 2|\mathcal{V}_i|$ sparse symmetric matrix. More explicitly, if $[v_{i_p},v_{i_q}] \subset \p \mathcal{S}_i$, we have 
\begin{equation}
M^{\text{area}}_{p,q+|\mathcal{V}|_i} = M^{\text{area}}_{q+|\mathcal{V}|_i,p} = 1 \ \text{ and } \ 
M^{\text{area}}_{q,p+|\mathcal{V}|_i} = M^{\text{area}}_{p+|\mathcal{V}|_i,q} = -1.
\end{equation}

DNCP minimizes the discrete conformal energy
\begin{equation}
E_C({\bf u}) = E_D({\bf u}) - A({\bf u})
\end{equation}
subject to the prescribed positions of two boundary vertices that remove the freedom of rigid motion and scaling. It suffices to solve a $2|\mathcal{V}_i| \times 2|\mathcal{V}_i|$ sparse linear system
\begin{equation}
\left(\begin{pmatrix}
L^{\text{cotan}} & 0 \\
0 & L^{\text{cotan}}
\end{pmatrix} - M^{\text{area}} \right) \begin{pmatrix} {\bf x} \\ {\bf y} \end{pmatrix} = 0
\end{equation}
subject to four boundary constraints (two in ${\bf x}$ and two in ${\bf y}$ for the two pinned boundary vertices). The resultant map $\varphi_i$ satisfying $\varphi_i(\mathcal{V}_i) = {\bf u} = [{\bf x}, {\bf y}]$ is the desired conformal parameterization of $\mathcal{S}_i$.

DNCP is suitable for our framework since it is a free-boundary linear method. As discussed above, obtaining each $\varphi_i$ only requires solving a $2|\mathcal{V}_i| \times 2|\mathcal{V}_i|$ sparse matrix equation, which is highly efficient. Also, the free-boundary condition ensures that no additional conformal distortion will be introduced at the boundaries. This is particularly important in our subsequent welding step.

It is noteworthy that the parameterization of each submesh is independent, and hence this step of computing local conformal parameterizations is highly parallelizable.

\subsection{Partial welding}
Note that the local parameterizations we obtained via DNCP are not necessarily consistent along the boundaries. Therefore, we need a step for gluing the boundaries of them consistently. To preserve the conformality of the parameterization, the gluing step should be conformal. This problem of gluing subdomains is different from the closed welding problem introduced in Section~\ref{sect:background}. More explicitly, the closed welding problem considers gluing the entire boundaries of two domains, while in general only a portion of the boundaries of two neighboring subdomains in our case should be glued. In other words, the problem that we need to tackle is a \emph{partial} welding problem that involves gluing two subdomains along only a pair of boundary arcs.

Below, we first rigorously derive a theoretical construction for solving the partial welding problem. Then, we devise an efficient algorithm for solving it.

\subsubsection{Theoretical construction}
We formulate the problem mathematically. Given two Jordan regions $A, B \subset \overline{\mathbb{C}}$, let $\gamma_A \subset \p A$ and $\gamma_B \subset \p B$ be some arcs of the boundaries of $A$ and $B$ respectively. Suppose we have a correspondence function $f: \gamma_A \to \gamma_B$ that relates points on $\gamma_A$ and points on $\gamma_B$. The partial welding problem is to find two conformal maps $\Phi_A: A \to A'$ and $\Phi_B: B \to B'$, with $A'$ and $B'$ being disjoint, such that
\begin{equation}
\Phi_A(\gamma_A) = (\Phi_B \circ f)(\gamma_A).
\end{equation}

Recall that the closed welding problem is solvable for quasisymmetric function on the real axis. For the partial welding problem, we make use of the following lemma.
\begin{lemma}[Lehto and Virtanen~\cite{Lehto73}] \label{lemma:lv}
Every function $f$ which is $k$-quasisymmetric on an interval $I = [a,b]$ can be extended to a $\tilde{k}$-quasisymmetric function on the entire $x$-axis, where the constant $\tilde{k}$ is less than a number depending only on $k$.
\end{lemma}

\begin{figure}[t]
\centering
\includegraphics[width=0.9\textwidth]{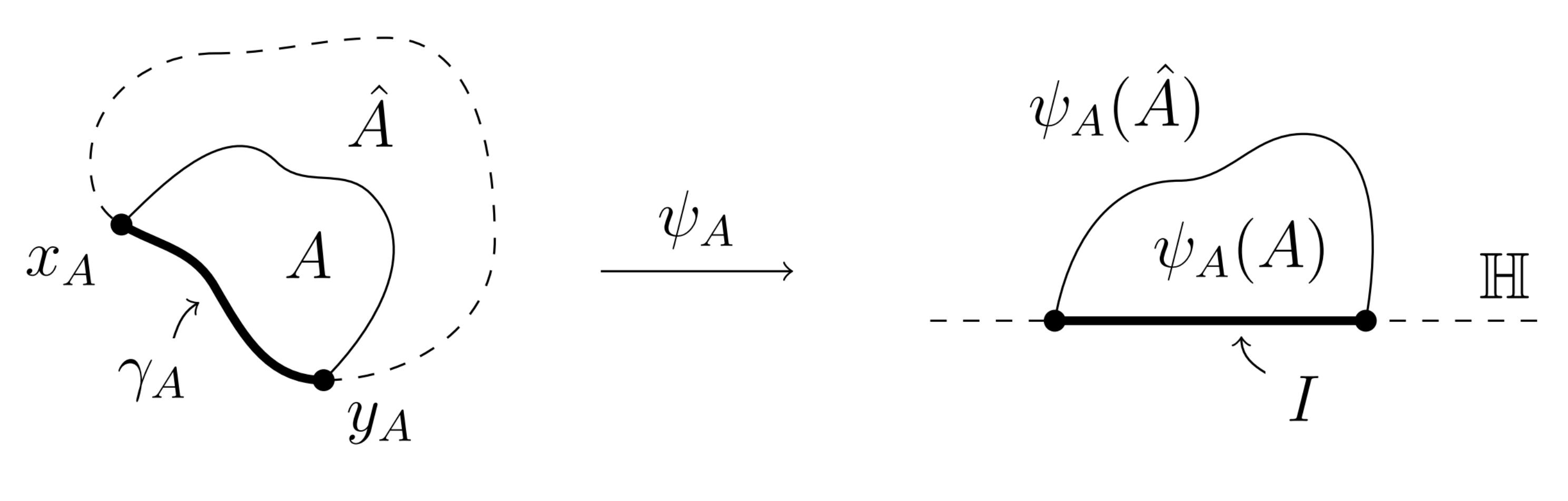}
\caption{Construction of $\hat{A}$ and the mapping from it to the upper half plane.}
\label{fig:Ahat_to_H}
\end{figure}

To make use of the above lemma, we further suppose that $\hat{A}$ is a larger domain that contains $A$ while sharing the boundary segment $\gamma_A$, i.e. $A \subset \hat{A}$ and $ \gamma_A \subset \p \hat{A}$ (see Figure~\ref{fig:Ahat_to_H} left). Denote the endpoints of $\gamma_A$ by $x_A$ and $y_A$. Similarly, let $\hat{B}$ be a domain such that $B \subset \hat{B}$ and $ \gamma_B \subset \p \hat{B}$, and denote the endpoints of $\gamma_B$ by $x_B$ and $y_B$.

By the Riemann mapping theorem, $\hat{A}$ and $\hat{B}$ can be mapped to the upper and lower half plane respectively by some conformal maps $\psi_A$ and $\psi_B$. Now, we fix $x_A$ and $y_A$ at the endpoints of some interval $I$ on the $x$-axis. For simplicity, we take $I = [-1,1]$ and fix $x_A$ and $y_A$ at $-1$ and $1$ respectively (see Figure~\ref{fig:Ahat_to_H} right). Similarly, we fix $x_B$ at $-1$ and $y_B$ at $1$. The homeomorphic extensions to the closures define a map $g: I \to I$ by $g = \psi_B \circ f \circ \psi_A^{-1}$. In other words, we have $f = \psi_B^{-1} \circ g \circ \psi_A$ by construction. Assuming that $g$ is a quasisymmetric function from $I$ to itself, we get a quasisymmetric extension $\hat{g}: \mathbb{R} \to \mathbb{R}$ of $g$ using Lemma~\ref{lemma:lv}. 

\begin{figure}[t]
\centering
\includegraphics[width=0.95\textwidth]{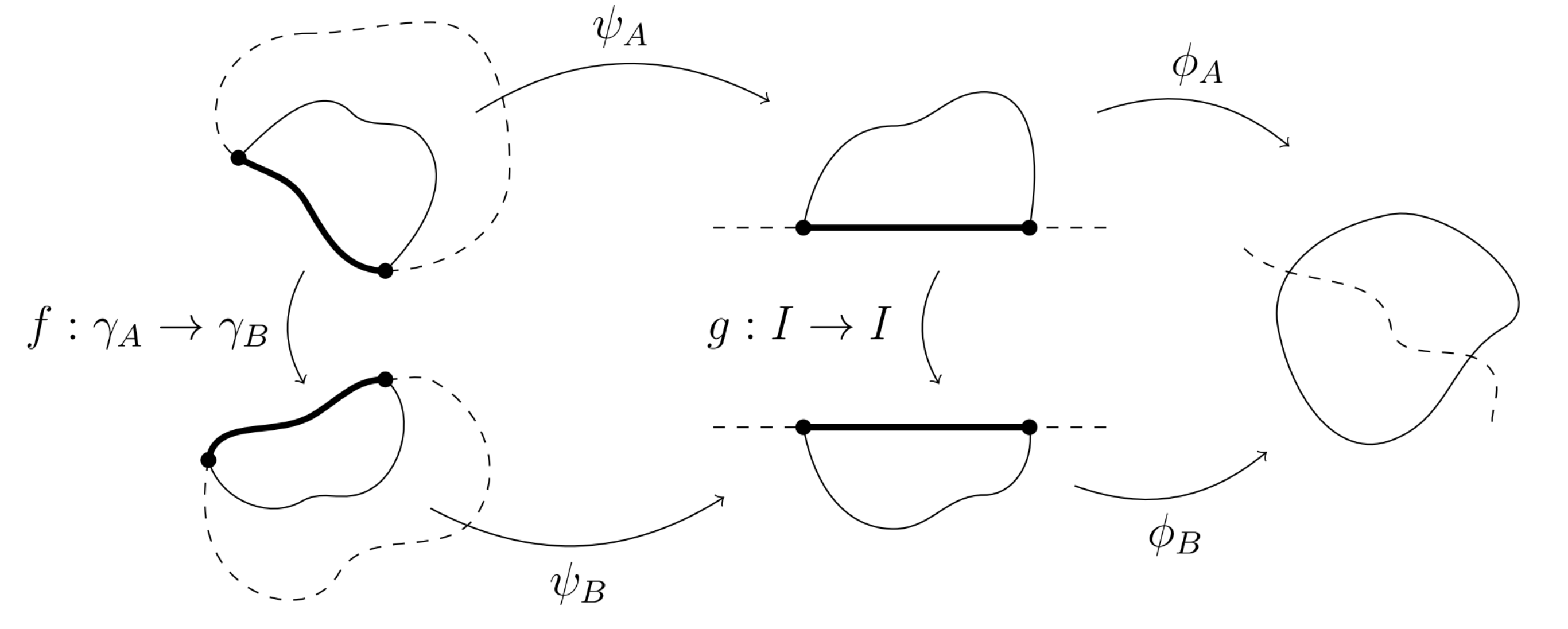}
\caption{The construction of conformal maps for solving the partial welding problem.}
\label{fig:quasisymmetric_extension}
\end{figure}

Then, we apply Theorem~\ref{thm:sewing} with this $\hat{g}$, which gives us two conformal maps $\phi_A: \hat{A} \to \hat{A}'$ and $\phi_B: \hat{B} \to \hat{B}'$ with $\hat{A}'$ and $\hat{B}'$ being disjoint, such that the boundary values satisfy $\phi_A(x) = \phi_B(\hat{g}(x))$ for all $x \in \mathbb{R}$. In particular, $\phi_A(x) = \phi_B(g(x))$ for all $x \in I$. Figure~\ref{fig:quasisymmetric_extension} shows an illustration of the construction.

Since the composition of conformal maps is conformal, we have constructed two conformal maps $\Phi_A = \phi_A \circ \psi_A$ and $\Phi_B = \phi_B \circ \psi_B$, which respectively map $A$ to some $A' \subset \hat{A}'$ and $B$ to some $B' \subset \hat{B}'$. Note that $f = \psi_B^{-1} \circ g \circ \psi_A$ when we restrict $f$ on $\gamma_A$. Also,
\begin{equation}
(\phi_B \circ \psi_B) \circ f = \phi_B \circ \psi_B \circ \psi_B^{-1} \circ g \circ \psi_A = \phi_B \circ g \circ \psi_A = \phi_A \circ \psi_A,
\end{equation}
where the last equality follows from Theorem~\ref{thm:sewing}. This solves the partial welding problem.

\subsubsection{Algorithmic construction}
The theoretical construction above provides us with a continuous approach for solving the partial welding problem. We proceed to develop an algorithm to solve the problem over discrete boundary data points. Suppose we have two sequences of boundary points $\p A = \{a_0, \dots, a_k, \dots, a_m\}$ and $\p B = \{b_0, \dots, b_k, \dots, b_n\}$, where $a_j$ corresponds to $b_j$ (i.e. $a_j$ should be glued with $b_j$) for $j = 0, \dots, k$. This gives a correspondence function $f: \gamma_A \subset \p A \to \gamma_B \subset \p B$, where $\gamma_A = \{a_0, \dots, a_k\}$ and $\gamma_B = \{b_0, \dots, b_k\}$, with $f(a_j) = b_j$ for $j = 0, \dots, k$. Our goal is to construct the two maps $\Phi_A, \Phi_B$ for gluing the two boundary curves conformally along the corresponding points. As discussed in the theoretical construction, a good way for the construction of $\Phi_A, \Phi_B$ is to make use of two maps $\psi_A, \psi_B$ that map $A,B$ onto the upper and lower half-planes respectively. We propose an algorithm that makes use of a variant of the geodesic algorithm~\cite{Marshall07}.

We begin with designing an algorithm that maps a sequence of boundary points to a standard shape. The algorithm is based on a key observation that the geodesic algorithm can be paused halfway. Suppose we have a sequence of boundary points $\{z_0, \dots, z_k, \dots, z_n\}$. Consider applying the first $k$ maps $g_1, g_2, \dots, g_k$ in the geodesic algorithm on $\{z_0, \dots, z_k, \dots, z_n\}$ with branching $(-1)^{1/2} = i$. The composition $g_k \circ \cdots \circ g_1$ maps the first $k+1$ points $\{z_0, \dots, z_k\}$ onto the imaginary axis, with $z_k$ mapped to 0, while the remaining boundary data points $\{z_{k+1}, \dots, z_n\}$ are all mapped onto the right half-plane. Note that each of $g_2, \dots, g_k$ is a composition of a M\"obius transformation, a square map and a square root map. Therefore, they are all conformal.

Now, instead of the final map \eqref{eqt:g_final_geodesic} in the geodesic algorithm, we apply the following M\"obius transformation:
\begin{equation}
g_{k+1}(z) = \frac{z}{1-\frac{z}{g_k \circ g_{k-1} \circ \cdots \circ g_1 (z_0)}}.
\end{equation}
It is easy to check that $g_{k+1}(g_k \circ \cdots \circ g_1 (z_0)) = \infty$ and $g(0) = 0$. In other words, the new composition $g_{k+1} \circ g_k \circ \cdots g_1$ maps the first data point $z_0$ to $\infty$ and the $(k+1)$-th data point $z_k$ to 0. Note that the first $k+1$ data points are on the upper half of the imaginary axis, and the remaining boundary data points $\{z_{k+1}, \dots, z_n\}$ are on the right half-plane. We call such a half-opened (i.e. half-unzipped) shape an \emph{intermediate form}. Note that by using another branching $(-1)^{1/2} = -i$ throughout the maps above, we have an alternative way to transform a sequence of boundary data points onto the right half-plane, with the first $k+1$ data points mapped onto the lower half of the imaginary axis. Algorithm~\ref{alg:intermediate_form} summarizes the proposed intermediate form transformation procedure.

\begin{algorithm2e}[h!]
\KwIn{A sequence of boundary points $\{z_0, \dots, z_k, \dots, z_n\}$ and a choice of branching.}
\KwOut{A sequence of transformed boundary points $\{Z_0, \dots, Z_k, \dots, Z_n\}$, where $Z_0, \dots, Z_k$ are on the imaginary axis.}
\BlankLine

Set $g_1(z) = \sqrt{\frac{z- z_1}{z-z_0}}$ with the choice of branching\;

\For{$j = 2, \dots, k$}{

Compute $\xi_j = (g_{j-1} \circ \cdots \circ g_1)(z_j)$\;

Set $g_j(z) = \sqrt{L_{\xi_j} (z)^2 -1}$ with the choice of branching, where $L_{\xi_j}(z) := \frac{\frac{\text{Re}(\xi_j)}{|\xi_j|^2}z}{1+\frac{\text{Im}(\xi_j)}{|\xi_j|^2}zi}$\;

}

Set $g_{k+1}(z) = \frac{z}{1-\frac{z}{g_k \circ g_{k-1} \circ \cdots \circ g_1 (z_0)}}$\;

Compute $P_l = (g_{k+1} \circ \cdots \circ g_1)(z_l)$ for $l = 0, \dots, k, \dots, n$\;

\caption{Intermediate form transformation}
\label{alg:intermediate_form}
\end{algorithm2e}

Coming back to the problem of aligning the two sequences of boundary points $\p A = \{a_0, \dots, a_k, \dots, a_m\}$ and $\p B = \{b_0, \dots, b_k, \dots, b_n\}$, we define auxiliary data points $a_{m+1} = b_{n+1} = 0, a_{m+2} = b_{n+2} =\infty$ to keep track of the transformation. Now, using Algorithm~\ref{alg:intermediate_form} with two different choices of branching $(-1)^{1/2} = i$ and $(-1)^{1/2} = -i$, we map $\{a_0, \dots, a_k, \dots, a_m, a_{m+1}, a_{m+2}\}$ and $\{b_0, \dots, b_k, \dots, b_n, b_{n+1}, b_{n+2}\}$ onto the right half-plane. Denote the transformed data points by $\{A_0, \dots, A_k, \dots, A_m, A_{m+1}, A_{m+2}\}$ and $\{B_0, \dots, B_k, \dots$ $, B_n, B_{n+1}, B_{n+2}\}$. Note that $\{A_0, \dots, A_k\}$ are all on the upper half of the imaginary axis with $A_0 = \infty$ and $A_k = 0$, while $\{B_0, \dots, B_k\}$ are all on the lower half of the imaginary axis with $B_0 = \infty$ and $B_k = 0$.  The next step is to align $A_j$ with $B_j$ for all $j = 0, \dots, k$ conformally, such that the two boundary curves $\p A$ and $\p B$ are welded based on the partial correspondence between $\gamma_A$ and $\gamma_B$. 

Suppose $\alpha = ai$ and $\beta = bi$ are two corresponding points originally on $\gamma_A$ and $\gamma_B$ under the intermediate form transformations, where $a>0>b$. A M\"obius transformation that takes $\{\alpha, 0, \beta\}$ to $\{i, 0, -i\}$ is explicitly given by
\begin{equation}
T_{\beta}^{\alpha}(z) = \frac{z}{\frac{-2ab}{a-b} - \frac{a+b}{a-b} zi}.
\end{equation}
This transformation provides us with a simple way to align each pair of corresponding points. Note that $A_k = 0 = B_k$ is automatically aligned, and so we start with aligning $A_{k-1}$ and $B_{k-1}$. Applying the M\"obius transformation $T_{B_{k-1}}^{A_{k-1}}$ onto the two sets of boundary data points, we map $A_{k-1}$ to $i$ and $B_{k-1}$ to $-i$. Then, we compose the map with the closing map $z \mapsto \sqrt{z^2 + 1}$ so that $i$ and $-i$ are both mapped to 0. More explicitly, we define
\begin{equation}
h_{k-1}(z):= \sqrt{T_{B_{k-1}}^{A_{k-1}}(z)^2 + 1}
\end{equation}
and apply it to all data points. The branching for the computation of each point is determined using the previous choice in the intermediate form transformation. Then, we repeat the above process for $j = k-2, \dots, 1$ by defining
\begin{equation}
h_j(z):= \sqrt{T_{\beta_j}^{\alpha_j}(z)^2 + 1},
\end{equation}
where
\begin{equation}
\alpha_j = (h_{j+1} \circ \cdots \circ h_{k-1})(A_j)
\end{equation}
and \begin{equation}
\beta_j = (h_{j+1} \circ \cdots \circ h_{k-1})(B_j).
\end{equation}

Now, all pairs of corresponding points $(A_1, B_1), \dots, (A_k, B_k)$ have been consistently aligned under the composition map $h_1 \circ h_2 \circ \dots \circ h_{k-1}$. The first pair of corresponding points $A_0 = \infty = B_0$ are also automatically aligned. Note that each of $h_{k-1}, \cdots, h_1$ is a composition of a M\"obius transformation, a square map and a square root map. Hence, they are all conformal. 

Then, we define a closing map $h_0$ similar to \eqref{eqt:g_final_geodesic} in the geodesic algorithm:
\begin{equation}
h_0(z):= \left(\frac{z}{1-\frac{z}{(h_{1} \circ \cdots \circ h_k)(\infty)}}\right)^2.
\end{equation}
Note that $h_0$ maps all points onto the upper half plane $\mathbb{H}$, with $0$ mapped to $0$ and $(h_{1} \circ \cdots \circ h_{k-1})(\infty)$ (i.e. $(h_{1} \circ \cdots \circ h_{k-1})(A_0)$) mapped to $\infty$. We obtain the transformed data points
\begin{equation}
\tilde{a}_l = (h_0 \circ \cdots \circ h_{k-1})(A_l)
\end{equation}
for $l = 0, \dots, m+2$ with branching $(-1)^{1/2} = i$, and 
\begin{equation}
\tilde{b}_l = (h_0 \circ \cdots \circ h_{k-1})(B_l)
\end{equation}
for $l = 0, \dots, n+2$ with branching $(-1)^{1/2} = -i$.

Considering the entire composition $h_0 \circ h_1 \circ \cdots h_{k-1} \circ g_{k+1} \circ \cdots g_1$ starting from the beginning to here, it can be observed that $g_2, \cdots, g_{k+1}, h_{k-1}, \cdots, h_2$ are all conformal, while $g_2$ is a square root map and $h_0$ is a square map. Therefore, the entire composition is conformal. In other words, we have conformally transformed the two sequences of boundary data points $\{a_0, \dots, a_k, \dots, a_m\}$ and $\{b_0, \dots, b_k, \dots, b_n\}$ into $\{\tilde{a}_0, \dots, \tilde{a}_k, \dots, \tilde{a}_m\}$ and $\{\tilde{b}_0, \dots, \tilde{b}_k, \dots , \tilde{b}_n\}$ such that the partial correspondence between them is satisfied, i.e. $\tilde{a}_j = \tilde{b}_j$ for $j = 0,\dots, k$.

Finally, we perform a normalization by tracking the transformation of the auxiliary data points $a_{m+1} = b_{n+1} = 0, a_{m+2} = b_{n+2} =\infty$. More explicitly, we apply a M\"obius transformation $T$ that takes $\{\tilde{a}_{m+1},\tilde{b}_{n+1},\frac{1}{2}(\tilde{a}_{m+2} + \tilde{a}_{n+2})\}$ to $\{-1,1,\infty\}$ on all the transformed points. This regularizes the transformation and prevents the boundary data points from being mapped far away. Note that M\"obius transformations are conformal and hence the conformality of the composition map is preserved. This completes the process of gluing two boundary curves based on a partial correspondence between them. Algorithm~\ref{alg:partial_welding} summarizes the proposed partial welding algorithm.

\begin{algorithm2e}[h]
\KwIn{Two sequences of boundary data points $\{a_0, \dots, a_k, \dots, a_m\}$ and $\{b_0, \dots, b_k, \dots, b_n\}$, where $a_j$ should be glued with $b_j$ for $j = 0, \dots, k$.}
\KwOut{Conformally transformed data points $\{\tilde{a}_0, \dots, \tilde{a}_k, \dots, \tilde{a}_m\}$ and $\{\tilde{b}_0, \dots, \tilde{b}_k, \dots , \tilde{b}_n\}$ such that $\tilde{a}_j = \tilde{b}_j$ for $j = 0,\dots, k$.}
\BlankLine

Define auxiliary points $a_{m+1} = b_{n+1} = 0, a_{m+2} = b_{n+2} =\infty$\;

Apply Algorithm~\ref{alg:intermediate_form} on $\{a_0, \dots, a_k, \dots, a_m, a_{m+1}, a_{m+2}\}$ with branching $(-1)^{1/2} = i$ and obtain the transformed boundary data points $\{A_0, \dots, A_k, \dots, A_m, A_{m+1}, A_{m+2}\}$\;

Apply Algorithm~\ref{alg:intermediate_form} on $\{b_0, \dots, b_k, \dots, b_n, b_{n+1}, b_{n+2}\}$ with branching $(-1)^{1/2} = -i$ and obtain the transformed boundary data points $\{B_0, \dots, B_k, \dots, B_n, B_{n+1}, B_{n+2}\}$\;

Set $h_{k-1}(z):= \sqrt{T_{B_{k-1}}^{A_{k-1}}(z)^2 + 1}$\;

\For{$j = k-2, \dots, 1$}{

Compute $\alpha_j = (h_{j+1} \circ \cdots \circ h_{k-1})(A_j)$ with branching $(-1)^{1/2} = i$\;

Compute $\beta_j = (h_{j+1} \circ \cdots \circ h_{k-1})(B_j)$ with branching $(-1)^{1/2} = -i$\;

Set $h_j(z):= \sqrt{T_{\beta_j}^{\alpha_j}(z)^2 + 1}$\;

}

Set $h_0(z):= \left(\frac{z}{1-\frac{z}{(h_{1} \circ \cdots \circ h_{k-1})(\infty)}}\right)^2$\;

Compute $\tilde{a}_l = (h_0 \circ \cdots \circ h_{k-1})(A_l)$ for $l = 0, \dots, m+2$, with branching $(-1)^{1/2} = i$\;

Compute $\tilde{b}_l = (h_0 \circ \cdots \circ h_{k-1})(B_l)$ for $l = 0, \dots, n+2$, with branching $(-1)^{1/2} = -i$\;

Apply a M\"obius transformation $T$ that takes $\{\tilde{a}_{m+1},\tilde{b}_{n+1},\frac{1}{2}(\tilde{a}_{m+2} + \tilde{a}_{n+2})\}$ to $\{-1,1,\infty\}$ on all the transformed points\;

\caption{Partial welding}
\label{alg:partial_welding}
\end{algorithm2e}

\begin{figure}[t!]
\centering
\includegraphics[width=\textwidth]{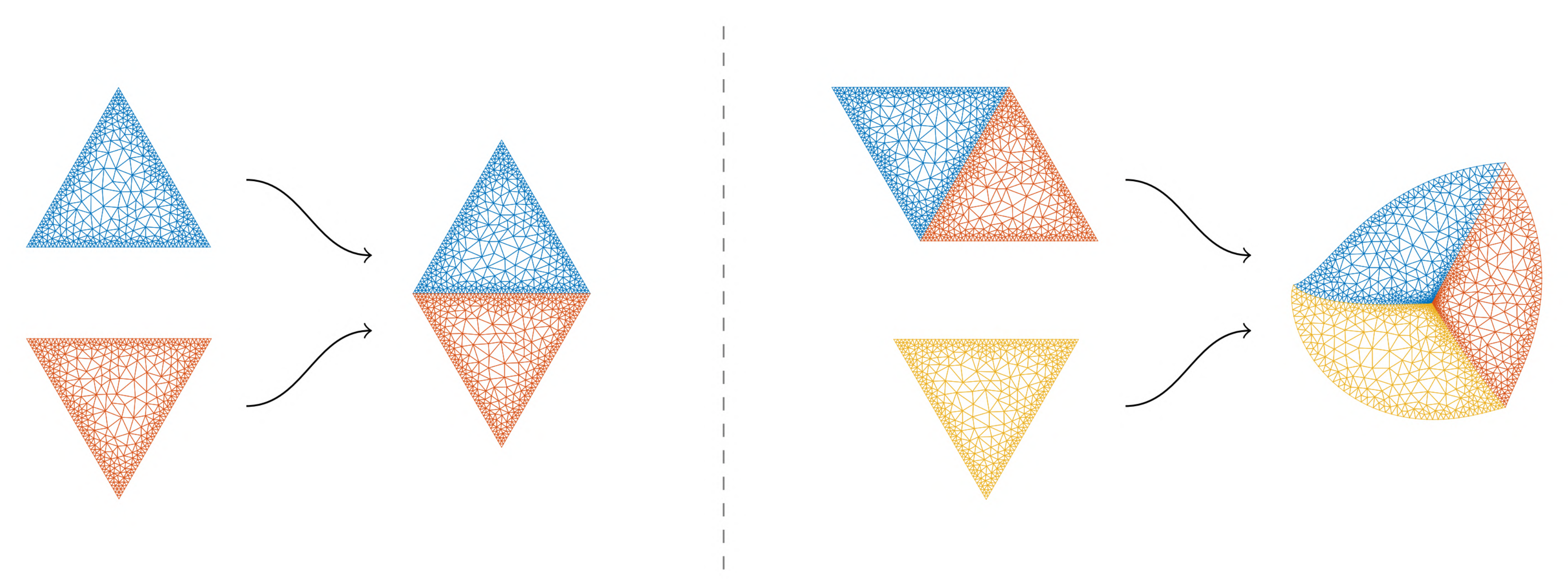}
\caption{An illustration of partial welding. Suppose we are given a pyramid-like surface with three triangular faces (colored in blue, red, yellow). Each of them have been flattened onto the plane. In order to weld the boundaries of the three triangles, we apply Algorithm~\ref{alg:partial_welding} twice. First, we glue the common boundaries of the blue and red triangles via partial welding. Then, we glue the common boundaries of the yellow triangle and the other two triangles via partial welding. Note that the interior points and the mesh structure of each triangle are plotted for a better visualization only. In the actual computation of partial welding, only the boundary points are involved.}
\label{fig:illustration_partial_welding}
\end{figure}

An illustration of the partial welding algorithm is given in Figure~\ref{fig:illustration_partial_welding}. As a remark, to weld two subdomains obtained by the local parameterization step partially, we only need to extract their boundary points on $\mathbb{C}$ and apply Algorithm~\ref{alg:partial_welding}. The interior points of the two flattened subdomains are not needed. With the updated coordinates of the boundary points of the subdomains, we can then easily obtain the desired global conformal parameterization by solving a number of sparse linear systems. The details will be described in Section~\ref{sect:main_last}.

\subsection{Enforcing additional constraints}
Before moving on to the step of obtaining the final global parameterization, it is possible for us to include an optional step here and enforce additional constraints for achieving disk conformal parameterization and spherical conformal parameterization.

\subsubsection{Constraints for disk conformal parameterization}
If the input simply-connected surface $\mathcal{S}$ is open, one can further restrict the target parameter domain to be the unit disk in the proposed method, thereby achieving a disk conformal parameterization. This is done by adding an extra step of applying the geodesic algorithm introduced in Section~\ref{sect:geodesic_algorithm} to the global boundary $\p \mathcal{S}$. Note that the points on $\p \mathcal{S}$ are distributed into various subdomains. Therefore, we first extract the coordinates of those boundary points from the partial welding result. Once the mapping that takes those points to the unit circle is determined, we apply the map for transforming the boundary coordinates of every flattened subdomain onto the unit disk. This results in boundary coordinates for the subdomains that yield a disk conformal parameterization upon solving the Laplace equation (details to be described in Section~\ref{sect:main_last}).

\subsubsection{Constraints for spherical conformal parameterization}
For genus-0 closed surfaces, one common choice of the parameter domain is the unit sphere $\mathbb{S}^2$. In case the input surface $\mathcal{S}$ is a genus-0 closed surface, we can modify our framework so that the partial welding procedure is repeated until two large components are left. Then, for the last welding, we use a closed welding instead of a partial welding to glue the entire boundaries of the two large components. As all boundaries are glued, the resulting boundary coordinates of the subdomains on the extended complex plane yield a spherical conformal parameterization upon solving the Laplace equation (details to be described in Section~\ref{sect:main_last}).

\subsection{Obtaining the global conformal parameterization}\label{sect:main_last}

After obtaining the new boundary constraints that satisfy the consistency condition, we can compute the global conformal parameterization of the input surface $\mathcal{S}$ by finding a harmonic map $\tilde{\varphi}_i: \mathcal{S}_i \to \mathbb{R}^2$ for each submesh with the new boundary constraints. More explicitly, it suffices to solve the Laplace equation 
\begin{equation}
\Delta \tilde{\varphi}_i = 0
\end{equation}
subject to the new boundary constraints. Again, note that the computations for the $K$ submeshes are independent and so this step is parallelizable. Because of the consistency between the boundaries of all subdomains, the new local parameterization results can be glued seamlessly, thereby forming a global conformal parameterization. One can further ensure the bijectivity of each subdomain using the idea of quasi-conformal composition (see Section~\ref{sect:qc}). More specifically, we compute the Beltrami coefficient of the inverse mapping ${\tilde{\varphi}_i}^{-1}$ (denoted by $\mu_{{\tilde{\varphi}_i}^{-1}}$). We can then determine whether $\tilde{\varphi}_i$ is folding-free by checking if $\|\mu_{{\tilde{\varphi}_i}^{-1}}\|_{\infty}> 1$ (or close to 1 in the discrete case). If so, we compose ${\tilde{\varphi}_i}$ with another mapping that is associated with the Beltrami coefficient $\mu_{{\tilde{\varphi}_i}^{-1}}$ to fix the fold-overs as guaranteed by quasi-conformal theory. With this additional step, we can ensure the bijectivity of the resulting global conformal parameterization.

Note that the resulting global parameterization lies in the extended complex plane. In case $\mathcal{S}$ is a genus-0 closed surface, we add a stereographic projection step to convert it to a spherical parameterization. Algorithm~\ref{alg:conformal_parameterization_open} summarizes the proposed method.

\begin{algorithm2e}[h!]
\label{alg:conformal_parameterization_open}
\KwIn{A simply-connected surface mesh $\mathcal{S} = (\mathcal{V},\mathcal{F})$, a set of edges $\tilde{\mathcal{E}}$ for the partition.}
\KwOut{A global conformal parameterization $\varphi:\mathcal{S} \to \mathbb{R}^2$ or $\mathbb{S}^2$.}
\BlankLine

Partition the mesh into $K$ submeshes based on $\tilde{\mathcal{E}}$\;

\For{$i = 1, \dots, K$}{
Compute a conformal parameterization of $\mathcal{S}_i = (\mathcal{V}_i,\mathcal{F}_i)$ using DNCP. Only the boundary coordinates of the parameterization are kept\;
}

Perform partial welding as described in Algorithm~\ref{alg:partial_welding} to update the boundary coordinates\;

(Optional) To achieve disk conformal parameterization, further apply the geodesic algorithm~\cite{Marshall07}. To achieve spherical conformal parameterization, perform conformal welding on the last two components obtained by partial welding\;

\For{$i = 1, \dots, K$}{

Solve the Laplace equation $\Delta \tilde{\varphi}_i = 0$ with the new boundary constraints for each $\mathcal{S}_i$\;
Compute the Beltrami coefficient $\mu_{{\tilde{\varphi}_i}^{-1}}$ to check whether $\tilde{\varphi}_i$ is folding-free. If not, fix the fold-overs in $\tilde{\varphi}_i$ using quasi-conformal composition\;

}

The solutions $\tilde{\varphi}_i$ for all $\mathcal{S}_i$ together form a global conformal parameterization $\varphi$. For spherical conformal parameterization, further apply the stenographic projection to map the result onto $\mathbb{S}^2$\;

\caption{Parallelizable global conformal parameterization of simply-connected surfaces (PGCP)}
\end{algorithm2e}

As a remark, the novel combination of local parameterization and partial welding in our proposed method significantly improves the computational efficiency of global conformal parameterization. Efficient sparse linear system solvers for the Laplace equation for the entire mesh ($2|\mathcal{V}| \times 2|\mathcal{V}|$) typically require a complexity of $O(|\mathcal{V}|^{1.5})$ (SOR), $O(|\mathcal{V}| \log |\mathcal{V}|)$ (FFT), $O(|\mathcal{V}|)$ (multigrid) etc.~\cite{Botsch05}. By contrast, one can see that the interior parts of the submeshes are not used in the partial welding step in our method. The partial welding step only involves $\mathcal{B}$, the collection of boundary points of the subdomains, with $|\mathcal{B}| \ll |\mathcal{V}|$. Also, the computation of the local parameterizations at the beginning and the harmonic maps at the end of our proposed method can both be parallelized, so that each computation involves $\mathcal{V}_i$ only.

\begin{figure}[t!]
\centering
\includegraphics[width=0.92\textwidth]{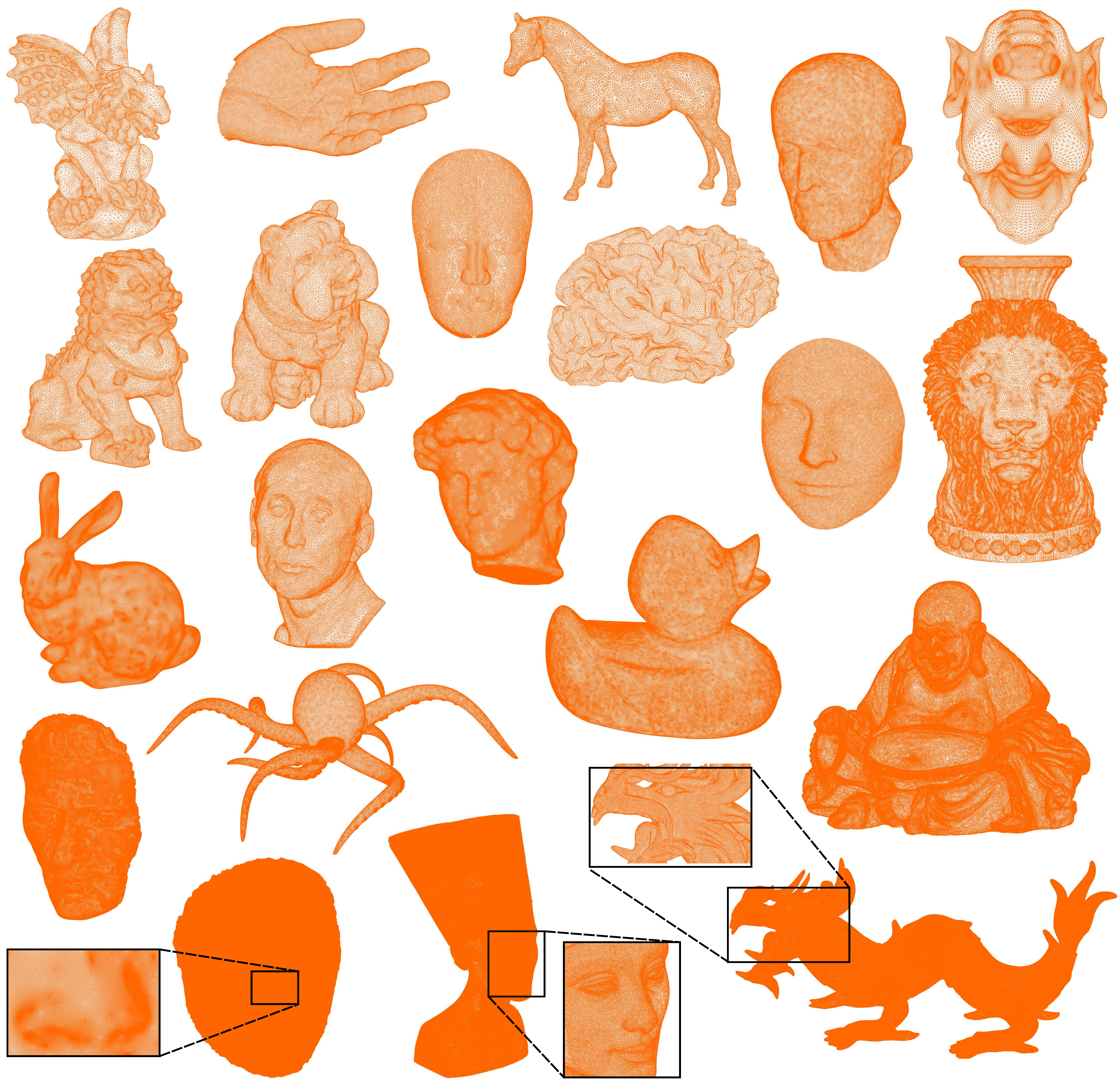}
\caption{A gallery of simply-connected surface meshes used in our experiments. Our PGCP method is capable of handling a wide range of simply-connected surfaces with different geometry, mesh quality and resolution.}
\label{fig:gallery}
\end{figure}

\section{Experiments}\label{sect:experiment}
Our proposed PGCP method is implemented in MATLAB, with the Parallel Computing Toolbox utilized for achieving parallelization. The sparse linear systems are solved using the backslash operator in MATLAB. All experiments in this section (except the experiment in Section \ref{sect:speedup}) are performed on a Windows PC with Intel i7-6700K quad-core CPU and 16~GB RAM. To evaluate the performance of our proposed method, we adapt various simply-connected surface meshes from multiple free 3D model repositories~\cite{aimatshape,Stanford,TurboSquid,Keenan} (see Figure~\ref{fig:gallery}). As for the distortion measure, we define the angular distortion of an angle $[v_i, v_j, v_k]$ (in degree) under the conformal parameterization $\varphi$ by
\begin{equation}
d([v_i, v_j, v_k]) = \angle [\varphi(v_i), \varphi(v_j), \varphi(v_k)] - \angle [v_i, v_j, v_k].
\end{equation}

\begin{figure}[t!]
\centering
\includegraphics[width=\textwidth]{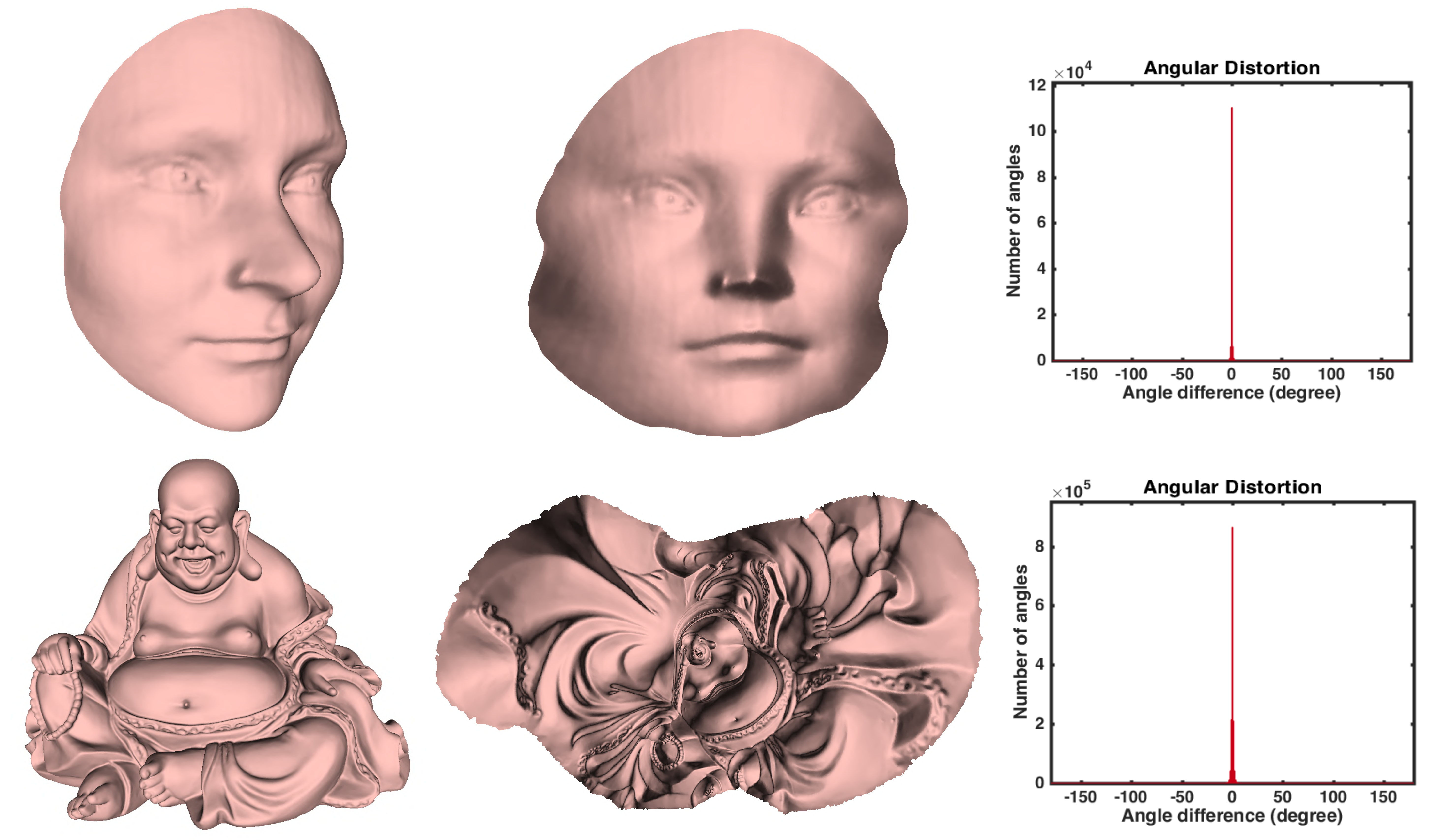}
\caption{Free-boundary conformal parameterizations of simply-connected open surfaces obtained by our proposed PGCP method, rendered with normal map shader.}
\label{fig:pgcp_free}
\end{figure}

\begin{table}[t!]
\small
\centering
\begin{tabular}{|C{20mm}|c|c|c|c|c|c|c|} \hline
\multirow{2}{*}{Surface} & \multirow{2}{*}{\# vertices} & \multicolumn{2}{c|}{SCP~\cite{Mullen08}} & \multicolumn{2}{c|}{CETM~\cite{Springborn08}} & \multicolumn{2}{c|}{PGCP} \\ \cline{3-8} 
& & Time (s) & mean($|d|$) & Time (s) & mean($|d|$) & Time (s) & mean($|d|$) \\ \hline 
Sophie & 21K & 1.1 & 0.2 & 1.2 & 0.2 & 0.6 & 0.2 \\ \hline
Niccol\`o da Uzzano & 25K & 1.3 & 0.6 & \multicolumn{2}{c|}{Failed} & 0.7 & 0.6  \\ \hline
Mask & 32K & 1.6 & 0.2 & 7.3 & 0.2 & 0.9 & 0.2  \\ \hline
Max Planck & 50K & 2.6 & 0.5 & 5.5 & 0.5 & 1.5 & 0.5 \\ \hline
Bunny & 85K &  4.4 & 0.5 & 18.8 & 0.5 & 2.0 & 0.5\\ \hline
Julius & 220K & 14.2 & 0.1 & 19.5 & 0.1 & 6.6 & 0.1 \\ \hline
Buddha & 240K & 13.7 & 0.6 & 49.0 & 0.6 & 9.2 & 0.6 \\ \hline
Face & 1M & 85.2 & $<0.1$ & 98.1 & $<0.1$ & 47.6 & $<0.1$ \\ \hline
\end{tabular}
\caption{The performance of spectral conformal parameterization (SCP)~\cite{Mullen08}, conformal equivalence of triangle meshes (CETM)~\cite{Springborn08} and PGCP for free-boundary conformal parameterization of simply-connected open surfaces.}
\label{table:free}
\end{table}

\subsection{Free-boundary conformal parameterization of simply-connected open surfaces}
We first consider computing free-boundary global conformal parameterization of simply-connected open surfaces using our proposed PGCP method (see Figure~\ref{fig:pgcp_free} for examples). To assess the performance of our method, we compare it with the spectral conformal parameterization (SCP)~\cite{Mullen08} and conformal equivalence of triangle meshes (CETM)~\cite{Springborn08} in terms of the computation time and the angular distortion (see Table~\ref{table:free}). The MATLAB version of SCP is implemented by the authors, and the MATLAB version of CETM can be found at~\cite{cetm}. The experimental results show that our proposed method is significantly faster than both SCP and CETM by over 40\% and 70\% respectively on average, while maintaining comparable accuracy in terms of the average angular distortion. This demonstrates the effectiveness of our method for free-boundary global conformal parameterization.

\subsection{Disk conformal parameterization of simply-connected open surfaces}
\begin{figure}[t!]
\centering
\includegraphics[width=\textwidth]{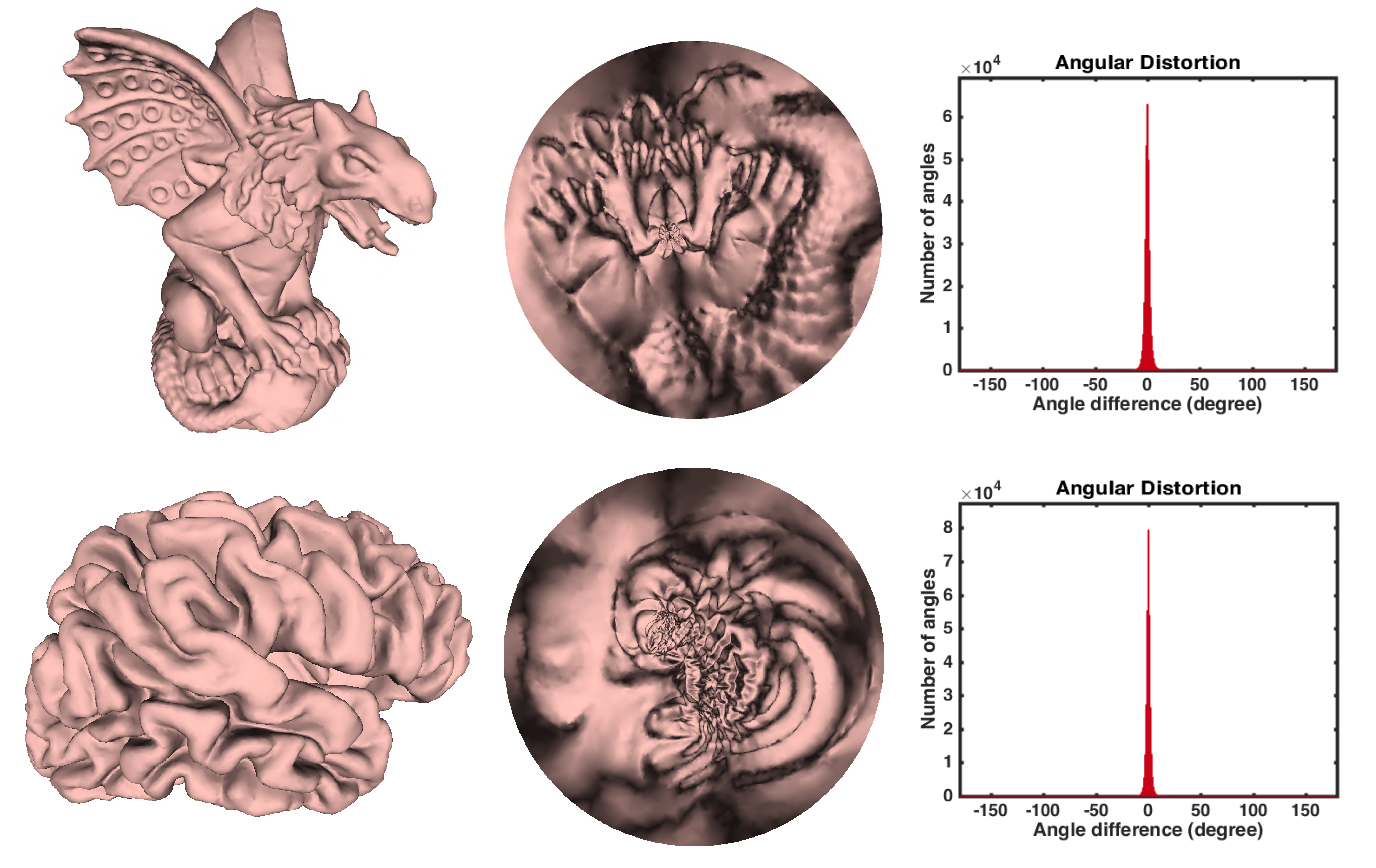}
\caption{Disk conformal parameterizations of simply-connected open surfaces obtained by our proposed PGCP method, rendered with normal map shader.}
\label{fig:pgcp_disk}
\end{figure}

Besides free-boundary global conformal parameterization, our proposed PGCP method can also achieve disk conformal parameterization of simply-connected open surfaces (see Figure~\ref{fig:pgcp_disk} for examples). To evaluate the performance of our method, we compare it with the state-of-the-art linear disk conformal map (LDM) method~\cite{Choi18a} and the conformal energy minimization (CEM) method~\cite{Yueh17} (see Table~\ref{table:disk}). The MATLAB version of LDM can be found at~\cite{ldm}, and the MATLAB version of CEM can be found at~\cite{cem}. It can be observed that our method is significantly faster than LDM and CEM by over 50\% and 30\% on average respectively. Also, our method achieves comparable or smaller angular distortion when compared to the two other methods. This shows that our method is advantageous for disk conformal parameterization.

\subsection{Spherical conformal parameterization of simply-connected closed surfaces}
\begin{figure}[t]
\centering
\includegraphics[width=\textwidth]{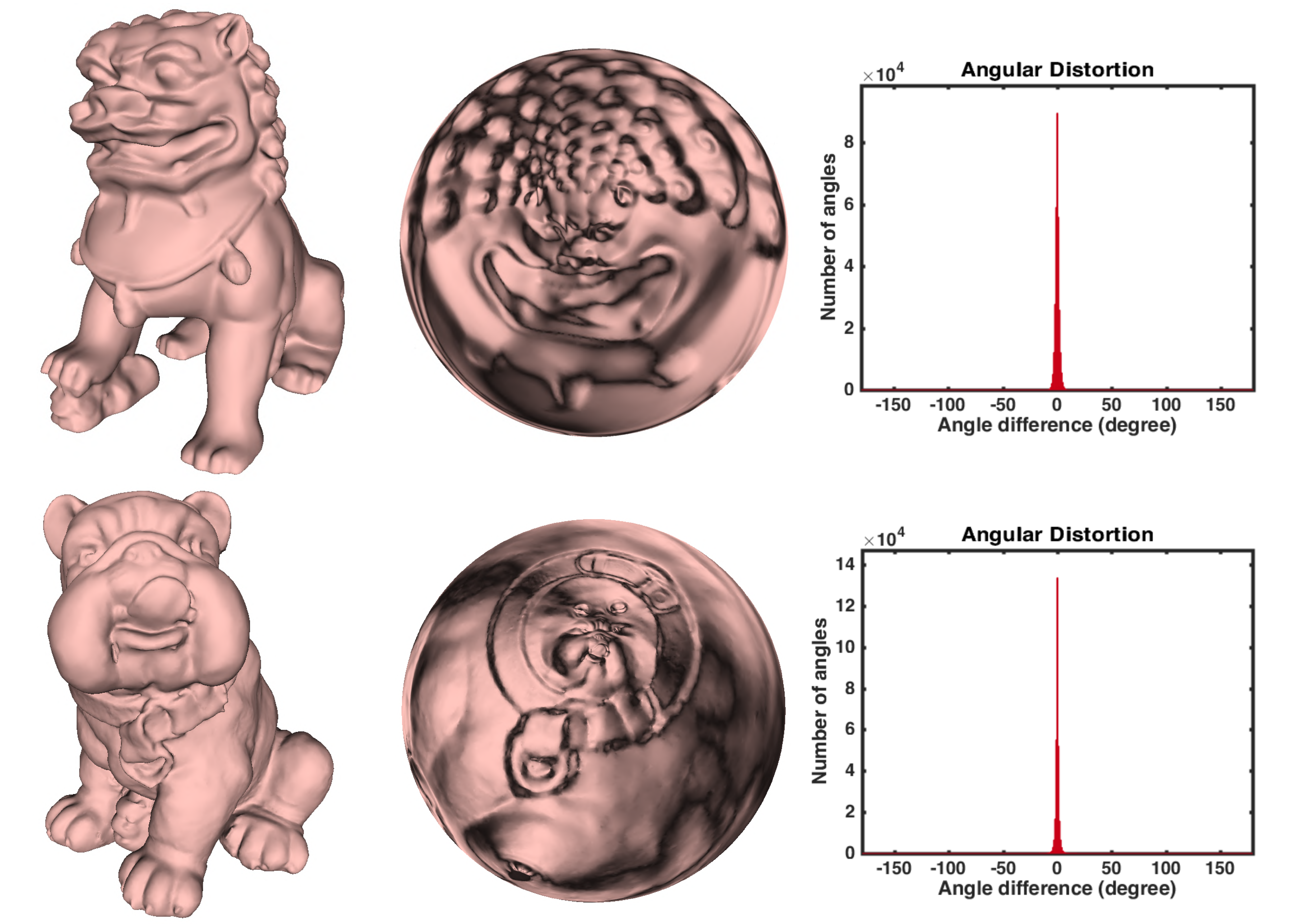}
\caption{Spherical conformal parameterizations of genus-0 closed surfaces obtained by our proposed PGCP method, rendered with normal map shader.}
\label{fig:pgcp_spherical}
\end{figure}

We then consider computing spherical conformal parameterization of genus-0 closed surfaces using our proposed PGCP method (see Figure~\ref{fig:pgcp_spherical} for examples). To evaluate the performance, we compare our proposed method with the state-of-the-art folding-free global conformal mapping (FFGCM) algorithm~\cite{Lai14} and the FLASH algorithm~\cite{Choi15a} (see Table~\ref{table:spherical}). The MATLAB version of FFGCM is kindly provided by its authors, and the MATLAB version of FLASH can be found at~\cite{flash}. Because of the ``divide-and-conquer'' nature of our method, our method is capable of producing spherical conformal parameterizations with a smaller angular distortion when compared to the two state-of-the-art algorithms. In particular, the FLASH algorithm involves puncturing a triangle from the input surface and flattening the punctured surface onto a big triangular domain. This step unavoidably creates squeezed regions and produces certain angular distortions. While the distortions are alleviated in the subsequent step using quasi-conformal composition, the step again involves a domain where most vertices are squeezed at the interior, which leads to some distortions. By contrast, our proposed PGCP method flattens each submesh naturally, with the shape of the submesh boundary taken into consideration. This effectively reduces the angular distortions, thereby producing a spherical conformal parameterization with a better accuracy. Moreover, because of the ability of exploiting parallelism, our method achieves a significant reduction in computational time by over 90\% on average when compared to FFGCM. When compared to FLASH, our method achieves comparable efficiency for moderate meshes and a notable reduction in computational time by around 25\% for dense meshes. This shows the advantages of our method for spherical conformal parameterization.

\begin{table}[t!]
\small
\centering
\begin{tabular}{|C{20mm}|c|c|c|c|c|c|c|} \hline
\multirow{2}{*}{Surface} & \multirow{2}{*}{\# vertices} & \multicolumn{2}{c|}{LDM~\cite{Choi18a}} & \multicolumn{2}{c|}{CEM~\cite{Yueh17}} & \multicolumn{2}{c|}{PGCP} \\ \cline{3-8} 
& & Time (s) & mean($|d|$) & Time (s) & mean($|d|$) & Time (s) & mean($|d|$) \\ \hline 
Ogre & 20K & 1.1 & 1.5 & 0.3 & 2.6 & 0.5 & 1.5\\ \hline
Niccol\`o da Uzzano & 25K & 1.6 & 0.8 & 1.4 & 1.3 & 0.8 & 0.8  \\ \hline
Brain & 48K & 2.9 & 1.6 & 2.9 & 1.5 & 1.3 & 1.5 \\ \hline
Gargoyle & 50K & 3.1 & 1.9 & 2.8 & 2.1 & 1.4 & 1.9 \\ \hline
Hand & 53K & 3.4 & 1.2 & 3.4 & 1.2 & 1.4 & 1.2 \\ \hline
Octopus & 150K & 15.4 & 7.2 & 10.4 & 24.0 & 8.9 & 5.6 \\ \hline
Buddha & 240K & 22.4 & 0.7 & 25.1 & 0.7 & 11.4 & 0.7 \\ \hline
Nefertiti & 1M & 87.9 & 2.9 & 83.2 & 4.2 & 52.7 & 2.9 \\ \hline 
\end{tabular}
\caption{The performance of linear disk conformal map (LDM)~\cite{Choi18a}, conformal energy minimization (CEM)~\cite{Yueh17} and PGCP for disk conformal parameterization of simply-connected open surfaces.}
\label{table:disk}
\end{table}

\begin{table}[t!]
\small
\centering
\begin{tabular}{|c|c|c|c|c|c|c|c|} \hline
\multirow{2}{*}{Surface} & \multirow{2}{*}{\# vertices} & \multicolumn{2}{c|}{FFGCM~\cite{Lai14}} & \multicolumn{2}{c|}{FLASH~\cite{Choi15a}} & \multicolumn{2}{c|}{PGCP} \\ \cline{3-8} 
& & Time (s) & mean($d$) & Time (s) & mean($d$) & Time (s) & mean($d$) \\ \hline 
Horse & 20K & 12.1 & 11.0 & 0.4 & 3.0 & 0.4 & 2.7 \\ \hline
Bulldog & 50K & 22.0 & 1.0 & 0.9 & 1.1 & 1.0 & 1.0  \\ \hline
Chinese Lion & 50K & 29.3 & 1.3 & 1.1 & 1.3 & 1.1 & 1.3 \\ \hline
Duck & 100K & 100.4 & 1.1 & 2.2 & 0.4 & 2.4 & 0.3 \\ \hline
David & 130K & 46.6 & 0.2 & 3.5 & 0.2 & 3.4 & 0.2 \\ \hline
Octopus & 150K & 112.3 & 37.2 & 10.1 & 6.9 & 7.1 & 2.6 \\ \hline
Lion Vase & 210K & 222.7 & 14.4 & 4.5 & 0.8 & 4.7 & 0.7 \\ \hline
Asian Dragon & 1M & \multicolumn{2}{c|}{Failed} & 64.4 & 1.3 & 48.5 & 0.9 \\ \hline
\end{tabular}
\caption{The performance of folding-free global conformal mapping (FFGCM)~\cite{Lai14}, FLASH~\cite{Choi15a} and PGCP for spherical conformal parameterization of genus-0 closed surfaces.}
\label{table:spherical}
\end{table}

\subsection{Comparison with boundary first flattening (BFF)}
While most of the existing methods can only handle a single type of global conformal parameterization, the recently proposed boundary first flattening (BFF) method~\cite{Sawhney18}, with code and executable files available at~\cite{bff}, is capable of computing multiple types of global conformal parameterizations, including free-boundary conformal parameterization for simply-connected open surfaces, disk conformal parameterization for simply-connected open surfaces, and spherical conformal parameterization for genus-0 closed surfaces. It is therefore natural to compare our proposed method and the BFF method.

\begin{table}[t!]
\small
\centering
\begin{tabular}{|c|C{20mm}|c|c|c|c|c|} \hline
\multirow{2}{*}{Parameterization} & \multirow{2}{*}{Surface} & \multirow{2}{*}{\# vertices} & \multicolumn{2}{c|}{BFF~\cite{Sawhney18}} & \multicolumn{2}{c|}{PGCP} \\ \cline{4-7} 
& & & Time (s) & mean($|d|$) & Time (s) & mean($|d|$) \\ \hline
\multirow{8}{*}{Free-boundary} & Sophie & 21K & 0.9 & 0.2 & 0.6 & 0.2 \\ \cline{2-7}
& Niccol\`o da Uzzano & 25K & 1.0 & 0.5 & 0.7 & 0.6  \\ \cline{2-7}
& Mask & 32K & 1.3 & 0.2  & 0.9 & 0.2  \\ \cline{2-7}
& Max Planck & 50K & 3.0 & 0.5  & 1.5 & 0.5 \\ \cline{2-7}
& Bunny & 85K &  4.7 & 1.2 & 2.0 & 0.5\\ \cline{2-7}
& Julius & 220K & 16.8 & 0.1 & 6.6 & 0.1 \\ \cline{2-7}
& Buddha & 240K & 14.4 & 0.6 & 9.2 & 0.6 \\ \cline{2-7}
& Face & 1M & \multicolumn{2}{c|}{Failed} & 47.6 & $<0.1$ \\ \hline

\multirow{8}{*}{Disk-boundary} & Ogre & 20K & 0.7 & 1.5  & 0.5 & 1.5\\ \cline{2-7}
& Niccol\`o da Uzzano & 25K & 1.0 & 1.1 & 0.8 & 0.8  \\ \cline{2-7}  
& Brain & 48K & 2.4 & 1.6 &  1.3 & 1.5 \\ \cline{2-7}
& Gargoyle & 50K & 2.1 & 1.9 & 1.4 & 1.9 \\ \cline{2-7} 
& Hand & 53K & 2.6 & 1.6  & 1.4 & 1.2 \\ \cline{2-7} 
& Octopus & 150K & 5.5 & 24.7 & 8.9 & 5.6 \\ \cline{2-7} 
& Buddha & 240K & 14.5 & 0.9 & 11.4 & 0.7 \\ \cline{2-7}  
& Nefertiti & 1M & \multicolumn{2}{c|}{Failed} & 52.7 & 2.9 \\ \hline %

\multirow{8}{*}{Spherical} & Horse & 20K & 1.0 & 74.52 & 0.4 & 2.7 \\ \cline{2-7} 
& Bulldog & 50K & 3.1 & 11.5 & 1.0 & 1.0  \\ \cline{2-7} 
& Chinese Lion & 50K & 3.0 & 4.4 & 1.1 & 1.3 \\ \cline{2-7} 
& Duck & 100K & 7.6 & 5.7 & 2.4 & 0.3 \\ \cline{2-7} 
& David & 130K & 15.9 & 2.6  & 3.4 & 0.2 \\ \cline{2-7} 
& Octopus & 150K & 10.2 & 74.3 & 7.1 & 2.6 \\ \cline{2-7} 
& Lion Vase & 210K & 14.2 & 4.9 & 4.7 & 0.7 \\ \cline{2-7} 
& Asian Dragon & 1M & \multicolumn{2}{c|}{Failed} & 48.5 & 0.9 \\ \hline

\end{tabular}
\caption{Comparison between BFF~\cite{Sawhney18} and PGCP for free-boundary conformal parameterization for simply-connected open surfaces, disk conformal parameterization for simply-connected open surfaces, and spherical conformal parameterization for genus-0 closed surfaces.}
\label{table:bff}
\end{table}

Table~\ref{table:bff} shows the comparison between the two methods. For free-boundary and disk conformal parameterization, it can be observed that our method achieves at least comparable and sometimes better conformality, with a shorter computational time. For spherical conformal parameterization, our method is advantageous in both the conformality and efficiency. A possible reason is that BFF handles genus-0 closed surfaces by removing an arbitrary vertex star, flattening the punctured surface onto a disk conformally followed by a suitable rescaling, and finally mapping the disk onto the sphere using stereographic projection and filling the punctured vertex star at the pole. The choice of the vertex star greatly affects the overall shape of the disk parameterization of the punctured surface and the final angular distortion of the spherical parameterization. By contrast, our ``divide and conquer'' approach enables us to tackle the conformal flattening problem of subdomains which are obtained from a more natural partition of the surface, thereby achieving better conformality.

\subsection{Robustness of the proposed parameterization method to the choice of cut edges}
\begin{figure}[t!]
\centering
\includegraphics[width=0.77\textwidth]{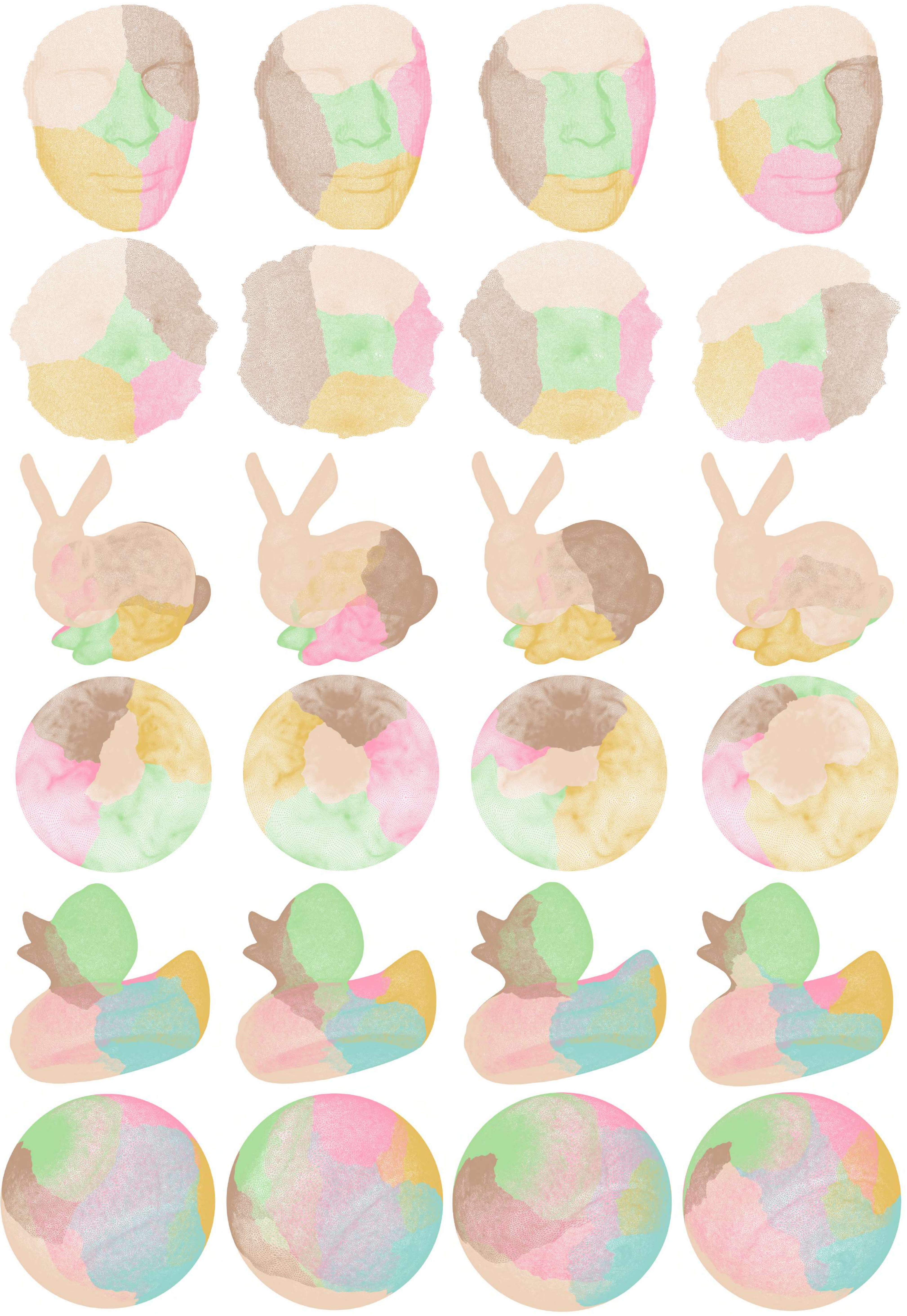}
\caption{The global conformal parameterizations produced by our PGCP method with different choices of cut paths. Each subdomain is with a distinct color. Top example: a simply-connected open human face surface with different cut paths, and the corresponding free-boundary conformal parameterization results. Middle example: a simply-connected open bunny surface with different cut paths, and the corresponding disk conformal parameterization results. Bottom example: a genus-0 closed duck surface with different cut paths, and the corresponding spherical conformal parameterization results.}
\label{fig:pgcp_robust}
\end{figure}

In our proposed PGCP method, there is a flexibility for the user to prescribe the cut paths and supply the set of chosen edges as an input of the parameterization algorithm. This allows the user to freely choose the number of subdomains to be used and how the surface is partitioned. It is natural to ask whether the performance of our method is robust to the choice of the cut edges. As shown in Figure~\ref{fig:pgcp_robust}, we consider different choices of cut paths on several surfaces and supplying them as an input of our proposed method. The accuracy of the resulting parameterizations is recorded in Table~\ref{table:cut_edges}. It can be observed that the angular distortions produced by different choices of cut paths are highly consistent, which indicates that our method is robust to the choice of the cut paths.

\begin{table}[t!]
\small
\centering
\begin{tabular}{|c|c|c|c|c|c|} \hline
Surface & Cut paths & mean($|d|$) & sd($|d|$) & median($|d|$) & iqr($|d|$)\\ \hline
\multirow{4}{*}{Face} & Figure~\ref{fig:pgcp_robust} (top, leftmost) & 0.25 & 0.44 & 0.13 & 0.23 \\  \cline{2-6} 
&  Figure~\ref{fig:pgcp_robust} (top, second left) & 0.26 & 0.44 & 0.16 & 0.24 \\ \cline{2-6}  
&  Figure~\ref{fig:pgcp_robust} (top, second right) & 0.24 & 0.44 & 0.12 & 0.24\\ \cline{2-6} 
&  Figure~\ref{fig:pgcp_robust} (top, rightmost) & 0.26 & 0.44 & 0.15 & 0.25\\ \hline  

\multirow{4}{*}{Bunny} & Figure~\ref{fig:pgcp_robust} (middle, leftmost) & 0.49 & 0.70 & 0.33 & 0.48 \\ \cline{2-6} 
&  Figure~\ref{fig:pgcp_robust} (middle, second left) & 0.49 & 0.70 & 0.33 & 0.48 \\ \cline{2-6}  
&  Figure~\ref{fig:pgcp_robust} (middle, second right) & 0.49 & 0.70 & 0.33 & 0.48\\ \cline{2-6} 
&  Figure~\ref{fig:pgcp_robust} (middle, rightmost) & 0.49 & 0.70 & 0.33 & 0.49\\ \hline  

\multirow{4}{*}{Duck} & Figure~\ref{fig:pgcp_robust} (bottom, leftmost) & 0.27 & 0.46 & 0.16 & 0.27\\ \cline{2-6} 
&  Figure~\ref{fig:pgcp_robust} (bottom, second left) & 0.27 & 0.46 & 0.16 & 0.27\\ \cline{2-6} 
&  Figure~\ref{fig:pgcp_robust} (bottom, second right) & 0.28 & 0.48 & 0.17 & 0.30\\ \cline{2-6} 
&  Figure~\ref{fig:pgcp_robust} (bottom, rightmost) & 0.27 & 0.47 & 0.16 & 0.27  \\ \hline 
\end{tabular}
\caption{The performance of our proposed PGCP method with different choices of cut paths. The mean, standard deviation, median, and interquartile range of the absolute angular distortion $|d|$ are evaluated.}
\label{table:cut_edges}
\end{table}

\subsection{Performance of our proposed PGCP method with different number of subdomains} \label{sect:speedup}

\begin{figure}[t]
\centering
\includegraphics[width=\textwidth]{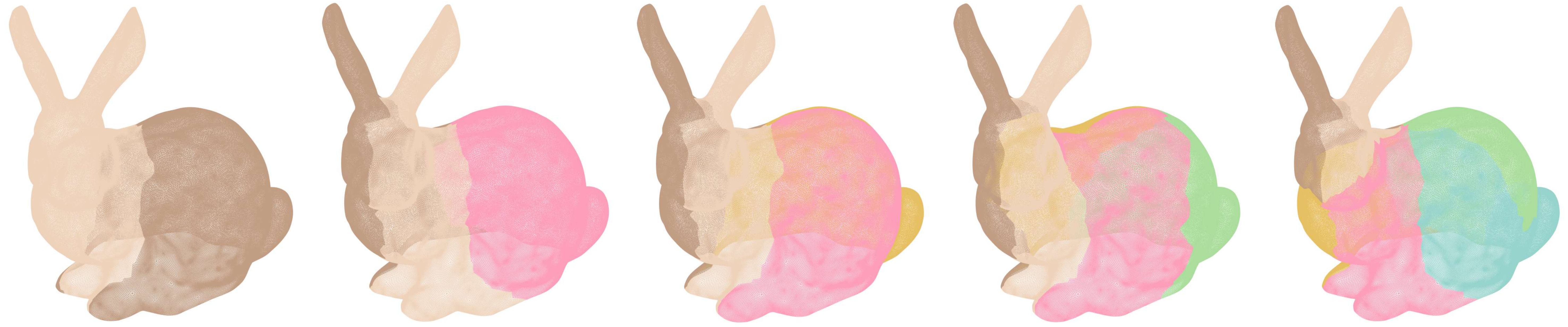}
\caption{The bunny model divided into different number of subdomains for the speedup experiment. Left to right: $2, 3, 4, 5, 6$.}
\label{fig:pgcp_speedup}
\end{figure}

By partitioning a given surface into $n$ subdomains evenly, one can reduce the original problem with size $|\mathcal{V}|$ to subproblems each with size $\frac{|\mathcal{V}|}{n}$. To check the parallel efficiency of our proposed method, we consider subdividing the bunny surface into different number of subdomains (see Figure \ref{fig:pgcp_speedup}) and apply our PGCP method for conformal parameterization. Since the maximum number of workers MATLAB creates depends on the number of physical cores, we use another machine with six physical cores for this experiment and focus on the ratio between the computational time with different number of subdomains. 

Table \ref{table:speedup} shows the experimental results with $2,3,4,5,6$ subdomains. Note that $2$ is used as the baseline as the welding algorithm requires at least two subdomains. From the experimental result, we  observe a speedup achieved by exploiting parallelization. To evaluate the parallel efficiency, we consider the ratio $E_n = \frac{S_n}{n/2}$, where $S_n$ is the speedup achieved by $n$ subdomains compared to the baseline and $n/2$ is the subdomain ratio. It can be observed that $E_n$ is close to 1 for small $n$ but shows a decreasing trend as $n$ increases. A possible reason is that in the numerical implementation, there are some technical steps which are not parallelizable and hence will cost additional time as the number of subdomains increases. For instance, both the step of partitioning the surface into subdomains and the step of combining the parameterization results of all subdomains as a large $|\mathcal{V}| \times 2$ or $|\mathcal{V}| \times 3$ matrix involve non-sliced variables (i.e. matrices that cannot be broken up into segments by MATLAB), such as \texttt{face(subdomain\_id == i,:)} where \texttt{face} is the original $|\mathcal{F}| \times 3$ triangulation, \texttt{subdomain\_id} is the ID of the connected component each triangle belongs to (after the mesh is cut along the set of cut edges $\mathcal{E}$). With the non-sliced variables, those computations are not parallelizable under the current MATLAB parallel computing (\texttt{parfor}) framework. For this reason, the ideal parallel efficiency cannot be achieved as $n$ increases. We anticipate that this issue can be alleviated by performing code optimization or considering alternative implementation languages.

\begin{table}[t!]
\small
\centering
\begin{tabular}{|c|c|c|c|c|c|} \hline
\# of subdomains $n$ & 2 & 3 & 4 & 5 & 6\\ \hline
Average mesh size $|\mathcal{V}|/n$ & 42K & 28K & 21K & 17K & 14K \\  \hline  
Parallel speedup $S_n$ & 1 & 1.31 & 1.66 & 1.91 & 2.05 \\ \hline   
$n/2$ & 1 & 1.5 & 2 & 2.5 & 3\\ \hline 
Parallel efficiency $E_n$ & 1 & 0.87 & 0.83 & 0.76 & 0.68 \\ \hline 
\end{tabular}
\caption{The performance of our proposed PGCP method for parameterizing the bunny model, with different number of subdomains used. Note that the welding algorithm requires at least two subdomains and hence we use $n=2$ as the baseline. For this reason, we define the parallel speedup to be $S_n = \frac{T_2}{T_n}$ where $T_i$ is the time taken with $i$ subdomains used, and the parallel efficiency to be $E_n = \frac{S_n}{n/2}$.}
\label{table:speedup}
\end{table}

\subsection{Applications}

\begin{figure}[t]
\centering
\includegraphics[width=\textwidth]{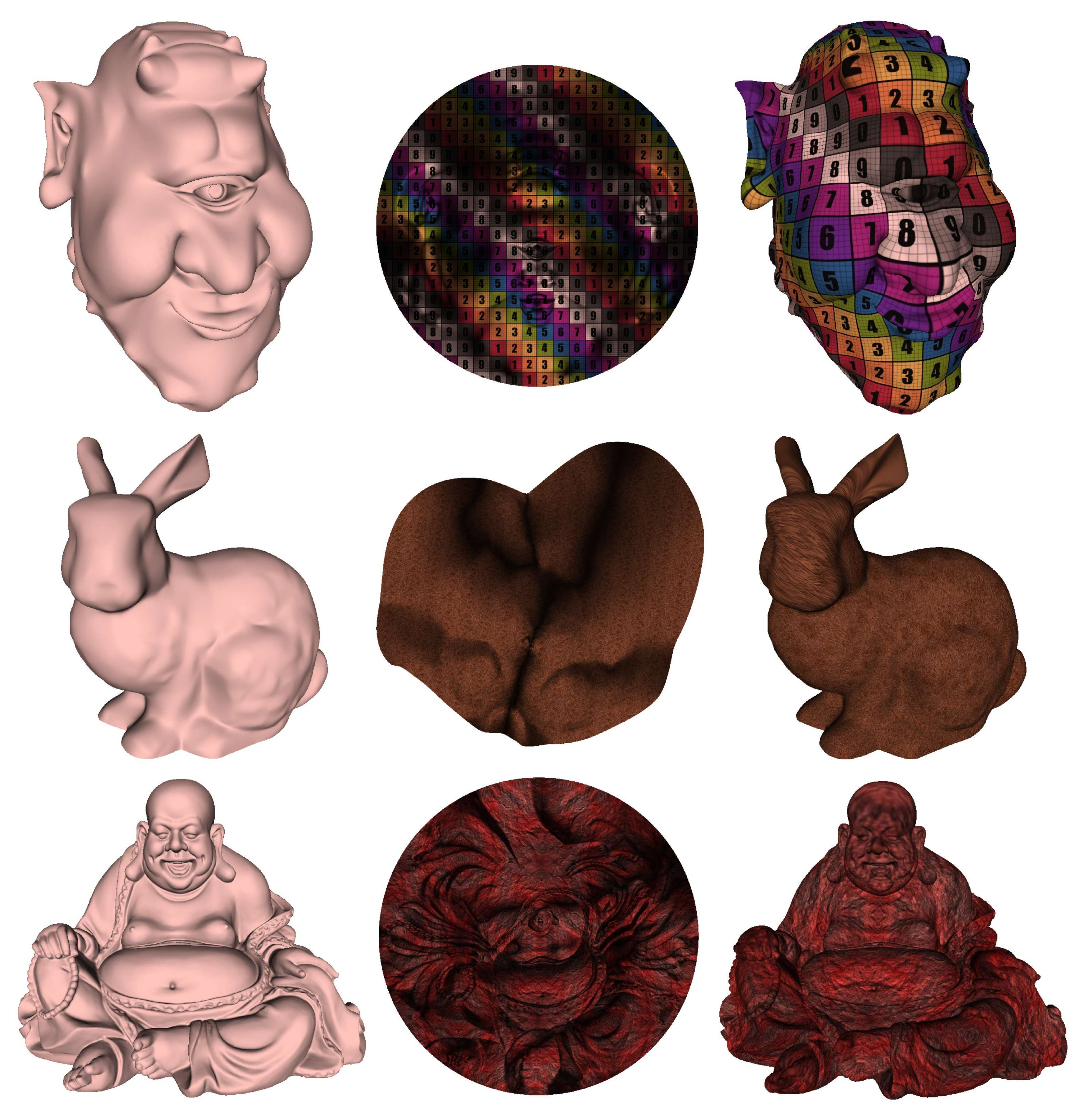}
\caption{Texture mapping via our proposed PGCP method. Left: the input mesh. Middle: the parameterization achieved by PGCP rendered with normal map shader, overlaid with a texture (colored checkerboard/hair/stone). Right: the texture mapping result.}
\label{fig:application_texture}
\end{figure}

The above experiments demonstrate the improvement of our proposed PGCP method over the state-of-the-art conformal parameterization algorithms. In this section, we discuss the applications of it.

\subsubsection{Texture mapping}
One application of our proposed PGCP method is texture mapping. After conformally flattening a surface onto the plane using our method, we can design a texture on the parameter domain. Since there is a 1-1 correspondence between the input surface and the parameter domain, we can then use the inverse mapping to map the texture back onto the surface, thereby obtaining a surface with the desired texture on it. Several examples are shown in Figure~\ref{fig:application_texture}. It is noteworthy that our method is conformal and hence the local geometry of the texture pattern is well preserved. For instance, the checkerboard texture shown in Figure~\ref{fig:application_texture} can maintain its orthogonality on the Ogre surface.

\subsubsection{Surface remeshing}
\begin{figure}[t]
\centering
\includegraphics[width=0.75\textwidth]{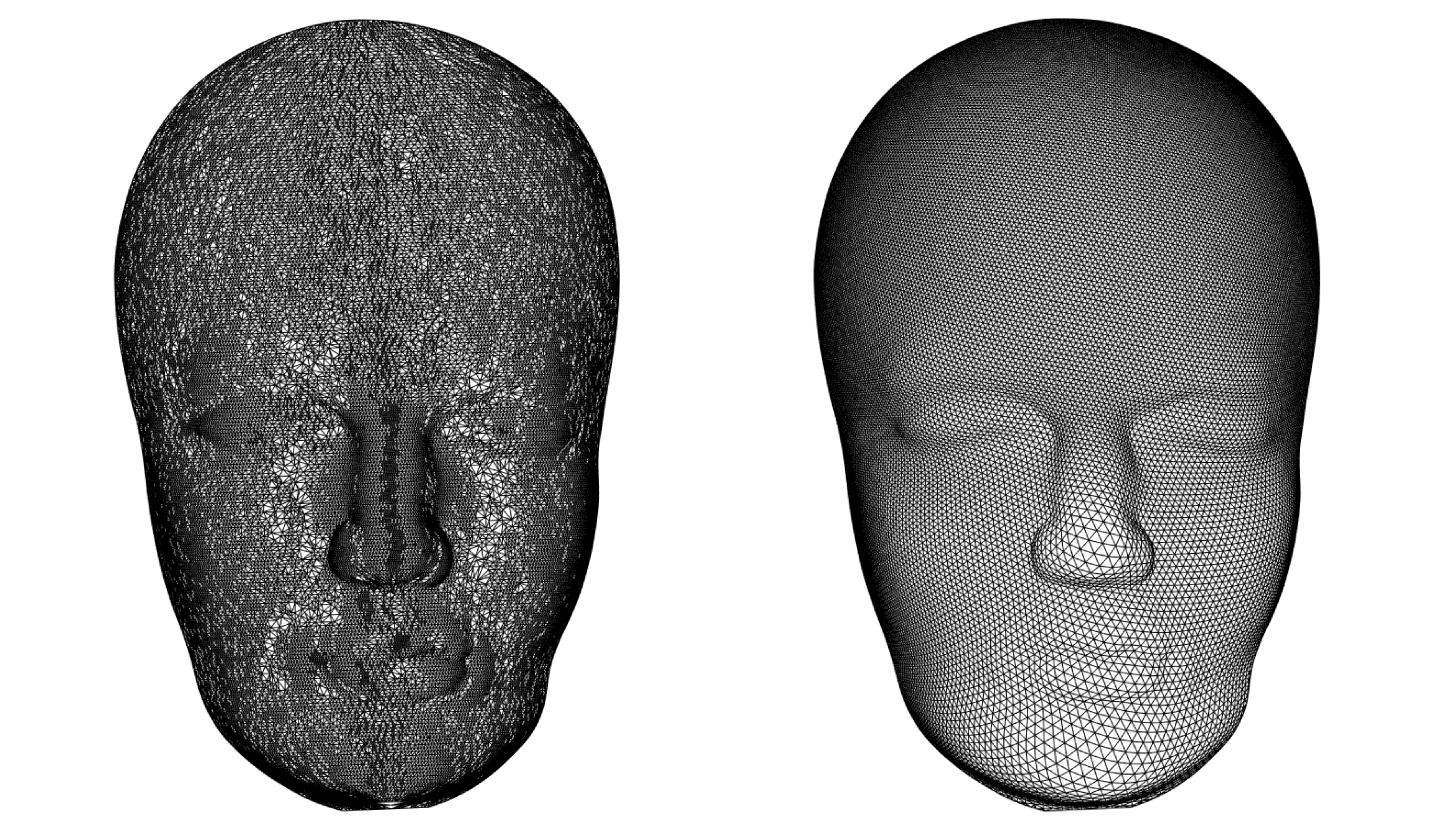}
\caption{Surface remeshing via our proposed PGCP method. Left: the input surface. Right: the remeshed surface with improved mesh quality.}
\label{fig:application_remeshing}
\end{figure}

Our proposed PGCP method can also be applied to surface remeshing, which aims at improving the mesh quality of a given surface. By conformally parameterizing the surface and constructing a regular mesh structure on the parameter domain, we can use the inverse mapping to map the mesh structure back onto the surface, thereby remeshing the surface (see Figure~\ref{fig:application_remeshing} for example). It is noteworthy that since the parameterization is conformal, the regularity of the mesh structure defined on the parameter domain is well-preserved on the surface. 

\subsubsection{Solving PDEs on surfaces}
\begin{figure}[t]
\centering
\includegraphics[width=\textwidth]{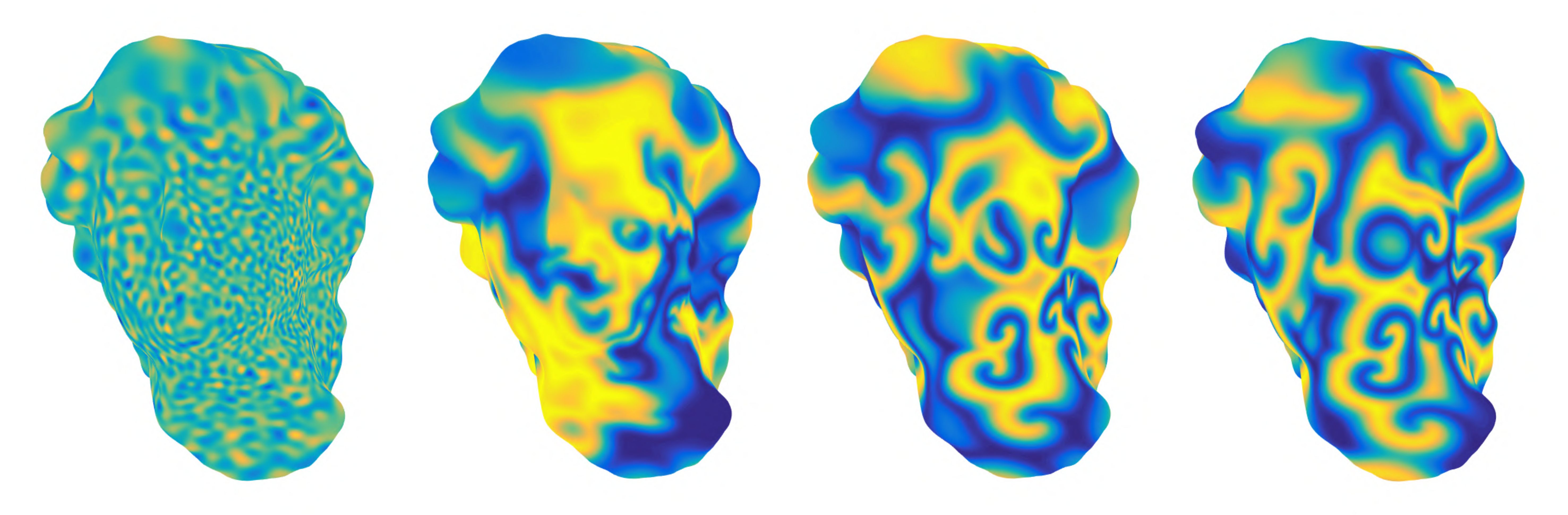}
\caption{Patterns formed on the genus-0 David surface by solving the Ginzburg-Landau equation on the spherical conformal parameterization. The leftmost is the initialization, and the rightmost is the final result.}
\label{fig:application_pde}
\end{figure}

Another notable application of our proposed PGCP method is solving PDEs on surfaces~\cite{Lui05}. While solving PDEs on a general surface is difficult, solving them on a standard parameter domain such as the unit sphere or the unit disk is relatively easy. Figure~\ref{fig:application_pde} shows an example of patterns formed on the genus-0 David surface by solving the time-dependent Ginzburg-Landau equation on the spherical conformal parameterization obtained by our proposed PGCP method. The PDE on the sphere is solved using Chebfun~\cite{Chebfun}. The example demonstrates the use of our method for PDE-based surface decoration. 

\subsubsection{Other applications}
Some other possible applications of conformal parameterizations include surface registration~\cite{Choi15a}, medical visualization~\cite{Choi17} and surface morphing~\cite{Yueh17}. As our proposed PGCP method is advantageous over the state-of-the-art algorithms in both the computational time and the conformal distortion, these tasks can be done with higher efficiency and accuracy using our method.

\section{Discussion} \label{sect:discussion}
\subsection{Conformality improvement}
From our experimental results as shown in Tables \ref{table:disk}, \ref{table:spherical}, \ref{table:bff}, it can be observed that the conformality improvement achieved by our method is particularly significant for meshes with elongated parts, such as the horse model and the octopus model (see Figure \ref{fig:gallery}). A possible explanation is that unlike the prior methods which compute the global conformal parameterization of a given surface by directly handling the entire surface, our method takes advantage of surface splitting. Note that in the step of solving the Laplace equation for the entire surface, the elongated parts of the surface are extremely squeezed relative to the boundary of the parameter domain (here the boundary corresponds to either the actual boundary of an open surface, or a triangle/vertex star of a closed surface), leading to numerical inaccuracy. Also, all elongated parts have to be taken into account in one single solve. By contrast, our method divides the meshes into subdomains and solves the Laplace equation for each subdomain, making the elongated parts less squeezed relative to the boundary of each flattened subdomain. Also, the number of elongated parts involved in each solve is reduced by the surface partition.  

\subsection{Area distortion}
Note that one common issue of global conformal parameterizations is that the area is largely distorted~\cite{Kharevych06}. While we have demonstrated the improvement in efficiency and conformality by our proposed method for computing global conformal parameterizations, one may also be interested in the area distortion produced by our method. To quantify the area distortion, for a surface mesh $\mathcal{S} = (\mathcal{V}, \mathcal{F})$ and a parameterization mapping $\varphi:\mathcal{S} \to \mathbb{R}^2$ or $\mathbb{S}^2$, we define the area distortion of a triangle $T \in \mathcal{F}$ by
\begin{equation}
d_{\text{area}}(T) = \log_e \frac{{\text{Area($f(T)$)}} / \left({\sum_{T' \in \mathcal{F}}\text{Area($f(T')$)}}\right)}{{\text{Area($T$)}} / \left({\sum_{T' \in \mathcal{F}}\text{Area($T'$)}}\right)}.
\end{equation}
In other words, $d_{\text{area}}$ measures the logged area ratio between the triangle in the original mesh and the corresponding triangle in the parameter domain, with two normalization factors removing the global area difference between the original mesh and the parameter domain. $d_{\text{area}} \approx 0$ indicates that the area distortion is small, and a large value of $|d_{\text{area}}|$ indicates that the area distortion is large (i.e. the triangle is either shrunk or magnified).

\begin{table}[t!]
\small
\centering
\begin{tabular}{|c|C{20mm}|c|c|C{25mm}|} \hline
Parameterization & Surface & BFF~\cite{Sawhney18} & PGCP & PGCP with an additional M\"obius transformation \\ \hline
\multirow{7}{*}{Free-boundary} & Sophie & 0.18 & 0.21 & 0.18 \\ \cline{2-5}  
& Niccol\`o da Uzzano &  0.60 & 0.84 & 0.57\\ \cline{2-5} 
& Mask & 0.24 & 0.58 & 0.32 \\ \cline{2-5} 
& Max Planck &  2.49 & 2.62 & 2.50 \\ \cline{2-5} 
& Bunny & 2.68 & 3.32 & 2.95 \\ \cline{2-5}  
& Julius & 0.28 & 1.04 & 0.52 \\ \cline{2-5}
& Buddha &  0.78 & 1.20 & 1.16\\ \hline 

\multirow{7}{*}{Disk-boundary} & Ogre & 1.21 & 1.21 & 1.21 \\ \cline{2-5}  
& Niccol\`o da Uzzano & 0.76 & 0.86 & 0.57 \\ \cline{2-5} 
& Brain & 2.17 & 2.13 & 2.13 \\ \cline{2-5} 
& Gargoyle & 3.90 & 3.90 & 3.87\\ \cline{2-5}  
& Hand& 5.29 & 5.25 & 5.25 \\ \cline{2-5}  
& Octopus & 6.79 & 8.13 & 8.13 \\ \cline{2-5}  
& Buddha & 0.78 & 0.79 & 0.77 \\ \hline  

\multirow{8}{*}{Spherical} & Horse & 27.03 & 8.90 & 6.54 \\ \cline{2-5} 
& Bulldog  & 6.74 & 1.09 & 1.08 \\  \cline{2-5} 
& Chinese Lion  & 4.46 & 1.93 & 1.74 \\  \cline{2-5}
& Duck  & 7.92 & 1.00 & 0.84 \\ \cline{2-5}  
& David  & 0.85 & 0.85 & 0.36 \\  \cline{2-5} 
& Octopus  & 26.95 & 26.44 & 26.19 \\  \cline{2-5} 
& Lion Vase  & 7.13 & 0.92 & 0.84 \\  \hline 

\end{tabular}
\caption{The area distortion mean($|d_{\text{area}}|$) of the global conformal parameterizations produced by the boundary first flattening (BFF) method~\cite{Sawhney18}, the proposed PGCP method, and the proposed PGCP method with an additional step of composing with a M\"obius transformation (\eqref{eqt:mobius_free} for free-boundary conformal parameterization, \eqref{eqt:mobius_disk} for disk-boundary conformal parameterization, and \eqref{eqt:mobius_free} together with the stereographic projection for spherical conformal parameterization).}
\label{table:area}
\end{table}

Table~\ref{table:area} shows the area distortion of the BFF method~\cite{Sawhney18} and the proposed PGCP method for various types of global conformal parameterizations. For spherical conformal parameterization, our method achieves a lower area distortion. For disk conformal parameterization, the two methods achieve similar area distortions. For free-boundary conformal parameterization, the BFF method possesses a lower area distortion.

The larger area distortion produced by our method in some cases is due to the lack of area control throughout the algorithm. To reduce the area distortion, one possible way is to include an extra step of composing the parameterization mapping $\varphi$ with some M\"obius transformations in our method. The conformality of the parameterization will be preserved as M\"obius transformations are conformal, and the area distortion can be  reduced by choosing a suitable M\"obius transformation. For instance, for free-boundary conformal parameterization, one can search for an optimal M\"obius transformation
\begin{equation} \label{eqt:mobius_free}
f(z) = \frac{az+b}{cz+d},
\end{equation}
where $a,b,c,d \in \mathbb{C}$ with $ad-bc \neq 0$, such that the composition $f \circ \varphi$ minimizes the area distortion. For disk conformal parameterization, one can similarly search for an optimal automorphism 
\begin{equation} \label{eqt:mobius_disk}
f(z) = \frac{z-\alpha}{1-\bar{\alpha} z},
\end{equation}
where $\alpha \in \mathbb{C}$ with $|\alpha|<1$, such that the composition $f \circ \varphi$ minimizes the area distortion. For spherical conformal parameterization, denote the stereographic projection by $\tau: \mathbb{S}^2 \to \overline{\mathbb{C}}$. One can search for an optimal M\"obius transformation as described in \eqref{eqt:mobius_free} such that the composition $\tau^{-1} \circ f \circ \tau \circ \varphi$ minimizes the area distortion. 

To verify this simple idea, we solve the above optimization problems using the MATLAB's optimization solver \texttt{fmincon}. For free-boundary and spherical conformal parameterizations, there are eight real parameters to be optimized (the real and imaginary parts of $a,b,c,d$). For disk conformal parameterizations, there are two real parameters to be optimized (the modulus and argument of $\alpha$). The final results are recorded in the rightmost column of Table~\ref{table:area}. It can be observed that the area distortion is effectively reduced with the aid of M\"obius transformations in many cases, especially for free-boundary conformal parameterizations and spherical conformal parameterizations. The improvement for disk conformal parameterizations is relatively less significant, possibly due to the smaller number of free parameters. To achieve an even lower area distortion may require the composition with some other conformal transformations, which we plan to explore in the future.

\subsection{Alternative numerical approaches for accelerating the computation}
While we have demonstrated the advantages of our proposed method over the prior conformal parameterization methods in terms of the computational time and conformality using the idea of partial welding, another path for accelerating the computation is to consider alternative numerical approaches for the prior methods. 

Note that the MATLAB code of CETM~\cite{Springborn08} solves an unconstrained optimization problem for the discrete conformal functional in each step using MATLAB's \texttt{fminunc} function, which is not parallelizable under the current MATLAB parallel computing framework (MATLAB only supports using parallel computing to estimate the numerical gradients for \texttt{fminunc} in case they are not supplied, but in the CETM code the gradients are already supplied). SCP~\cite{Mullen08} involves solving a generalized eigenvalue problem which is done in MATLAB using the \texttt{eigs} function. However, multithreading is currently not supported for this function for sparse matrices. Replacing these MATLAB functions by external routines is not straightforward and so we proceed to consider accelerating the CEM method~\cite{Yueh17} for disk conformal parameterizations. Note that CEM is an iterative method that solves two Laplace equations (one for the boundary nodes and one for the interior nodes) for each step, and the Laplacian matrices are unchanged throughout the iterations. Here we consider combining CEM with the Combinatorial Multigrid (CMG) method~\cite{koutis2011combinatorial}, which is a hybrid graph-theoretic algebraic multigrid solver that combines the strengths of multigrid with those of combinatorial preconditioning, with MATLAB implementation publicly available~\cite{cmg}. We replace the backslash solve ($\setminus$) in the MATLAB implementation of CEM~\cite{cem} by the CMG solver. Table \ref{table:disk_alternative} shows the performance of CEM, CEM combined with CMG, and our proposed method. It can be observed that while CEM combined with CMG demonstrates an improvement in efficiency for large problems when compared to the original CEM, our proposed method is still more advantageous in terms of both the efficiency and conformality.

There are many other numerical approaches and solvers which are worth exploring as alternative paths toward parallelization, such as the parallel sparse linear system solvers by Koutis and Miller~\cite{Koutis07}, Peng and Spielman~\cite{Peng14} and Kyng {\it et al.}~\cite{Kyng16}, the \emph{Lean Algebraic Multigrid (LAMG)} method~\cite{Livne12} for solving the sparse linear systems involved in the prior parameterization methods. Incorporating these alternative numerical approaches into the current codes will require a careful consideration of the linkers and the cost of communication between different software and external libraries. Moreover, even if exploiting parallelization for solving the sparse linear systems can speed up the computation of the prior conformal parameterization methods, unlike our proposed partial welding approach, this approach is unable to improve their conformality or allow them to handle a wider class of surfaces. 

\begin{table}[t!]
\small
\centering
\begin{tabular}{|C{20mm}|c|c|c|c|c|c|c|} \hline
\multirow{2}{*}{Surface} & \multirow{2}{*}{\# vertices} & \multicolumn{2}{c|}{CEM~\cite{Yueh17}} & \multicolumn{2}{c|}{CEM~\cite{Yueh17} with CMG} & \multicolumn{2}{c|}{PGCP} \\ \cline{3-8} 
& & Time (s) & mean($|d|$) & Time (s) & mean($|d|$) & Time (s) & mean($|d|$) \\ \hline 
Ogre & 20K  & 0.3 & 2.6 & 0.3 & 2.6 & 0.5 & 1.5\\ \hline
Niccol\`o da Uzzano & 25K & 1.4 & 1.3 & 1.5 & 1.4 & 0.8 & 0.8  \\ \hline
Brain & 48K & 2.9 & 1.5 & 2.8 & 1.5 & 1.3 & 1.5 \\ \hline
Gargoyle & 50K & 2.8 & 2.1 & 3.2 & 2.1 & 1.4 & 1.9 \\ \hline
Hand & 53K & 3.4 & 1.2 & 3.2 & 1.4 & 1.4 & 1.2 \\ \hline
Octopus & 150K & 10.4 & 24.0 & 9.3 & 26.6 & 8.9 & 5.6 \\ \hline
Buddha & 240K  & 25.1 & 0.7 & 18.5 & 0.9 & 11.4 & 0.7 \\ \hline
Nefertiti & 1M & 83.2 & 4.2 & 74.4 & 4.2 & 52.7 & 2.9 \\ \hline 
\end{tabular}
\caption{The performance of the conformal energy minimization (CEM)~\cite{Yueh17} method, CEM combined with the Combinatorial Multigrid (CMG)~\cite{koutis2011combinatorial}, and our proposed PGCP method for disk conformal parameterization of simply-connected open surfaces.}
\label{table:disk_alternative}
\end{table}

\section{Conclusion}\label{sect:conclusion}
In this work, we have proposed a novel parallelizable global conformal parameterization method called PGCP for simply-connected surfaces. Given a triangle mesh, we partition it into submeshes and conformally flatten each of them using DNCP. As the local parameterization results do not yield a consistent global parameterization, we extract their boundary points to integrate them using a novel technique called partial welding. Using the modified boundaries for all submeshes, harmonic maps can be computed to yield a global conformal parameterization, with bijectivity guaranteed by quasi-conformal theory. Additional steps can be included to produce disk conformal parameterizations for surfaces with boundary, and spherical conformal parameterizations for genus-0 closed surfaces.

Most parts of our proposed method, such as the initial local conformal parameterization step and the last harmonic mapping step, can be computed independently in a distributed manner. The only global computation involved in our algorithm takes merely boundary data points of the submeshes, which are much fewer than the vertices of the entire mesh. Experimental results have demonstrated the significant improvement in efficiency and accuracy achieved by our proposed method when compared to the state-of-the-art approaches for free-boundary conformal parameterization, disk conformal parameterization and spherical conformal parameterization.

For future work, we plan to explore the possibility of extending our method for quasi-conformal parameterizations and mappings~\cite{Choi16b,Choi18c,Choi19a}. More specifically, note that the partial welding step in our proposed method is conformal, and the quasi-conformal dilatation of a map is preserved under the composition with conformal maps. Therefore, it should be possible for us to compute quasi-conformal parameterizations and mappings for dense meshes by a combination of local quasi-conformal maps of submeshes and partial welding. Another possible future work is the extension of our method for point clouds. As the partial welding approach uses only the boundary data points of the flattened submeshes but not the mesh structure of them, it should also be applicable for subdomains of a point cloud. Combining the partial welding approach with some existing conformal parameterization methods for disk-type point clouds will then yield a parallelizable global conformal parameterization method for point clouds.

\end{document}